\documentclass[pdflatex,sn-jnl,Numbered,icol]{sn-jnl}

\usepackage{graphicx}%
\usepackage{multirow}%
\usepackage{amsmath,amssymb,amsfonts}%
\usepackage{amsthm}%
\usepackage{mathrsfs}%
\usepackage[title]{appendix}%
\usepackage[table]{xcolor}%
\usepackage{textcomp}%
\usepackage{manyfoot}%
\usepackage{booktabs}%
\usepackage{algorithm}%
\usepackage{algorithmicx}%
\usepackage{algpseudocode}%
\usepackage{listings}%
\usepackage{caption}
\usepackage{amsmath}
\usepackage{array}
\usepackage{bm}
\usepackage{subcaption}
\usepackage{siunitx}
\usepackage{changes}
\usepackage{natbib}

\renewcommand{\added}[1]{#1}
\renewcommand{\deleted}[1]{}
\renewcommand{\replaced}[2]{#1}

\newcounter{change}
\newcommand{\chglabel}[1]{\refstepcounter{change}\label{#1}}

\begin{document}

\title[Article Title]{Towards a Real-Time Simulation of Elastoplastic Deformation Using Multi-Task Neural Networks}


\author[1]{\fnm{R.} \sur{Schmeitz}}\email{r.schmeitz@tue.nl }

\author[1,2]{\fnm{J.J.C.} \sur{Remmers}}\email{ j.j.c.remmers@tue.nl}

\author[3]{\fnm{O.} \sur{Mula}}\email{ o.mula@tue.nl}

\author*[1]{\fnm{O.} \sur{van der Sluis}}\email{ o.v.d.sluis@tue.nl}

\affil[1]{\small\orgdiv{Department Mechanical Engineering}, \orgname{Eindhoven University of Technology}, \orgaddress{\city{Eindhoven}, \country{The Netherlands}}}

\affil[2]{\small \orgname{Eindhoven AI Systems Institute, \orgaddress{\street{Laplace 32}}, \postcode{5612 AJ}, \city{Eindhoven}, \country{The Netherlands}}}

\affil[3]{\small\orgdiv{Department Mathematics and Computer Science}, \orgname{Eindhoven University of Technology}, \orgaddress{\city{Eindhoven}, \country{The Netherlands}}}

\abstract{
This study introduces a surrogate modeling framework merging proper orthogonal decomposition, long short-term memory networks, and multi-task learning, to accurately predict elastoplastic deformations in real-time. Superior to single-task neural networks, this approach achieves a mean absolute error below 0.40\% across various state variables, with the multi-task model showing enhanced generalization by mitigating overfitting through shared layers. Moreover, in our use cases, a pre-trained multi-task model can effectively train additional variables with as few as 20 samples, demonstrating its deep understanding of complex scenarios. This is notably efficient compared to single-task models, which typically require around 100 samples.

Significantly faster than traditional finite element analysis, our model accelerates computations by approximately a million times, making it a substantial advancement for real-time predictive modeling in engineering applications. While it necessitates further testing on more intricate models, this framework shows substantial promise in elevating both efficiency and accuracy in engineering applications, particularly for real-time scenarios.}


\maketitle

\section{Introduction}\label{sec1} 
In various technical and medical domains accurately modeling complex path-dependent deformations of materials presents a significant computational challenge \citep{Legrand2021ASurgery}. This is due to the nonlinearity of materials and complex geometries. The finite element method (FEM) is a widely recognized approach in solid mechanics, renowned for its ability to address boundary value problems systematically \citep{Okereke2018ComputationalMethod}. However, as spatial discretization is refined to enhance accuracy, the computational demands of FEM increase significantly. The challenges of path-dependent material non-linearity and the resultant limitations on practicality for real-time applications underscore the importance of achieving real-time predictive capabilities. Such capabilities are vital for critical applications where a Digital Twin is indispensable, including surgical assistance (for preoperative planning and virtual training systems), and autonomous vehicles (for navigation and obstacle avoidance) \citep{Marinkovic2019SurveySimulations, Morooka2008Real-TimeDeformation, Huber2018HighlyAspects, Xiong2022DesignScenario}. 

In response to these limitations, reduction techniques, such as Proper Orthogonal Decomposition (POD), have emerged as valuable solutions \citep{Sanghi2011ProperApplications}. These methods effectively reduce the dimensionality of full-order data, leading to lower computational costs while maintaining high accuracy \citep{Soldner2017AHomogenisation}. For linear boundary value problems, projection-based reduction methods offer an efficient means of reducing costs without compromising accuracy significantly. Unfortunately, this approach is less efficient for nonlinear systems that necessitate repetitive computations, owing to the requirement for constructing system matrices and carrying out inverse calculations. This complexity is further amplified when history-dependent material behavior is involved; in such cases, solving boundary value problems becomes computationally expensive due to the need for extensive calculations across all loading paths. To address these challenges, hyperreduction methods have been developed \citep{Jain2019Hyper-ReductionSystems}. These methods aim to alleviate the computational costs associated with nonlinear terms in large systems by strategically selecting a small set of nodes (or elements) in the mesh over which the nonlinearity is evaluated \citep{Cho2020EnhancedProblems}. This approach employs a combination of interpolation and extrapolation methods to accurately approximate the nonlinearity across the entire mesh, effectively lowering computational demands.

Despite progress in reduced-order modeling methods, nonlinear systems still present substantial computational challenges \citep{Bond2007ASystems, Rewienski2003ADevices}. Machine learning (ML) models, particularly neural networks (NN's), are emerging as a potent solution to these challenges. By optimizing neuron connections during training, these neural networks are proficient in capturing the linear correlations between inputs and outputs \citep{Lopez-Pacheco2022ComplexModeling}. A crucial element in enhancing their capability lies in the integration of activation functions, which are instrumental in empowering models to accurately represent nonlinear functions \citep{Leshno1993MultilayerFunction}. Utilizing a trained neural network in this way helps to bypass computational limitations commonly encountered with conventional methods. This method addresses the difficulties present in highly nonlinear analyses, such as those found in various fluid dynamics applications \citep{Brunton2021ApplyingMechanics}. 

\added{Building on these strengths, recurrent neural networks (RNNs) have emerged as powerful tool, particularly long short-term memory (LSTM) networks, effectively model complex temporal dependencies, such as strain-stress responses \citep{Zhang2021ApplicationSoil}. This temporal modeling capability is particularly valuable for accurately simulating material behavior over time, addressing a core challenge in structural and material analysis.}\chglabel{1.16a}

In the field of solid mechanics, \citet{Im2021SurrogateDecomposition} recently used POD-assisted LSTM networks to model elastoplastic deformations. Their approach trains separate NN's to predict different state variables from POD data for a specific problem. While showing accurate prediction, their methodology lacks a shared representation between tasks and requires training individual models for each problem setup. This limitation implies a potential inefficiency in adapting the method to varied problem settings without extensive retraining, highlighting the need for more versatile and adaptive modeling approaches.

\begin{figure}[H]
  \centering
  \includegraphics[width=0.8\textwidth]{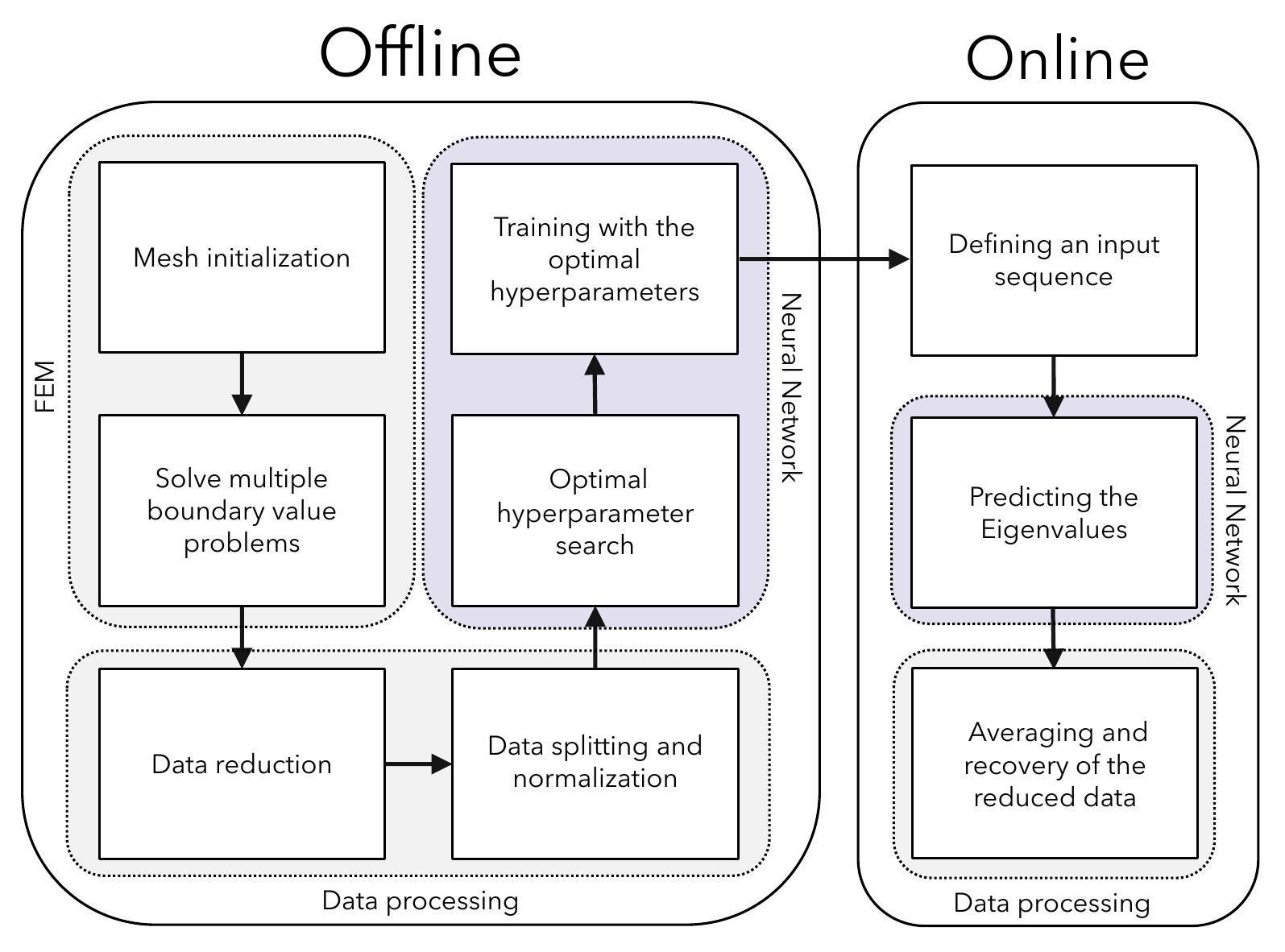}\captionsetup{justification=centering} 
  \caption{Framework.}
  \label{fig:Process}
\end{figure}

\added{This study is motivated by the potential of multi-task learning (MTL) neural network architectures. MTL has gained traction within different fields such as solid mechanics \citep{Diao2023SolvingTechnology, Liu2021DeepNetwork} due to its ability to enhance predictive accuracy by training on multiple related tasks simultaneously.} \chglabel{1.16b} By sharing learned features across tasks, MTL frameworks support complex behavior modeling, enabling efficiency improvements in multi-output problems like stress-strain interactions and displacement fields. 
\added{
This flexibility enables the model to be trained on additional parameters relevant to specific applications, such as surgical assistance. As new requirements arise, such as predicting material behavior or integrating operational parameters, the pre-trained model can be easily adjusted without the need for complete retraining. This approach enhances efficiency and contributes to a more sustainable modeling process by reducing resource consumption and training time.}\chglabel{2.2ea} This adaptability reduces the number of samples required for effective training and significantly decreases computation time, leading to a more sustainable solution. 
Incorporating MTL with RNNs can further improve versatility and generalization in complex material behavior modeling.
The proposed framework, unlike traditional single-task models, merges these shared and specific layers. It surpasses previous reduction methods' limitations and balances computational speed with accuracy for real-time uses. Illustrated in Figure \ref{fig:Process}, our framework integrates machine learning with traditional techniques such as FEM, aiding in complex material behavior modeling and enhancing real-time prediction accuracy. This paper aims to demonstrate our framework's ability to learn universal representations for precise predictions across diverse field variables, for elastoplastic material behavior.
\vspace{-0.3cm}
To achieve this goal, the paper first explores the used methods, including the finite element method, elastoplasticity, data reduction, and neural networks. Building on this basis, the LSTM multi-task framework is presented, elucidating its architecture, training methodology, and evaluation metrics. The framework is validated through an in-depth study of two distinct 2D scenarios, focusing on small strain elastoplasticity. This involves a comprehensive review of the setup, data gathering, data reduction, and the results. Following the confirmation of the framework's effectiveness, the performance of the multi-task architecture is assessed. The paper concludes by identifying limitations and suggesting avenues for future research.

\section{Hardening Plasticity Model}\label{sec2}
To construct a precise and efficient surrogate model with neural networks, it is critical to first generate training data by solving boundary value problems (BVP) through finite element simulations. This step ensures a rich dataset that reflects the complex dynamics. The assumptions and considerations of this model, are elaborated within this section.

\subsection{Model formulation}
We address a BVP designed for elastoplastic materials, specifically incorporating kinematic hardening to precisely model material behavior under various loading conditions.

\begin{figure}[H]
\centering
\includegraphics[width=0.5\textwidth]{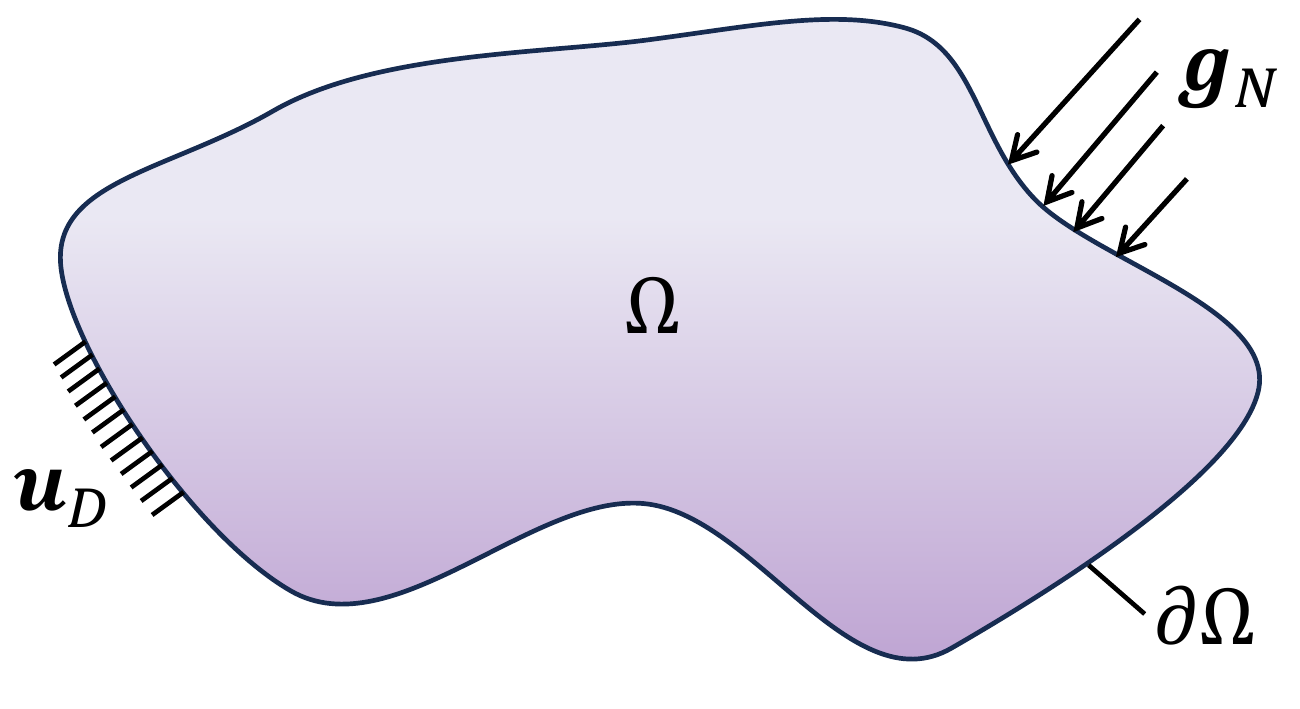}
\captionsetup{justification=centering} 
\caption{Domain of the Boundary Value Problem with its corresponding boundary conditions.}
\label{fig:BVP}
\end{figure}
\noindent
For the boundary-value problem under consideration, we specify a continuum body through its domain \( \Omega \subset \mathbb{R}^d \), with \( d \in \{2,3\} \) indicating the spatial dimensionality of the system. This body is examined in its reference configuration and analyzed over a finite time interval \( [0, T) \) where \( T > 0 \). The body is subject to boundary conditions on its perimeter, \( \partial \Omega \), divided into a Dirichlet boundary, \( \partial \Omega_D \), where fixed displacements \( \boldsymbol{u}_D \) are prescribed, and a Neumann boundary, \( \partial \Omega_N \), upon which a normal force \( \boldsymbol{g}_N:\Omega\to \mathbb{R}^d \) is imposed. These conditions are represented as:
\[
\boldsymbol{u} = \boldsymbol{u}_D \text{ on } [0,T) \times \partial\Omega_D, \quad \boldsymbol{\sigma} \cdot \boldsymbol{n} = \boldsymbol{g}_N \text{ on } [0,T) \times \partial\Omega_N.
\]
\noindent
Note that the prescribed displacements \( \boldsymbol{u}_D \) and the applied normal force \( \boldsymbol{g}_N \) can be time-dependent.

The primary objective is to determine the displacement field \( \boldsymbol{u}: [0, T) \times \Omega \to \mathbb{R}^d \), that results from the applied boundary conditions, neglecting any body force effects. Following these specifications, the equilibrium equation in its strong form is:

\begin{equation}
\nabla \cdot \boldsymbol{\sigma} = 0.
\end{equation}
\\
\noindent
Here, \( \boldsymbol{\sigma} \in \mathbb{R}_\text{sym}^{d \times d} \) represents the stress tensor, which is symmetric. Assuming a small strain formulation \( \boldsymbol{\varepsilon} := \frac{1}{2} \left( \nabla \boldsymbol{u} + (\nabla \boldsymbol{u})^\top \right) \), the constitutive relation for a linear elastoplastic material is expressed as:
\noindent
\begin{equation}
    \boldsymbol{\sigma} := \boldsymbol{\mathbb{C}} : (\boldsymbol{\varepsilon}-\boldsymbol{\varepsilon}_{pl}),
\end{equation}
\noindent
where \( \boldsymbol{\mathbb{C}}  \) denotes the fourth-order elasticity tensor, which is determined by the material's Young's modulus \( E \) and Poisson's ratio \( \nu \). The term \( \boldsymbol{\varepsilon}_{pl} \in \mathbb{R}_\text{sym}^{d \times d} \) denotes the plastic strain tensor.

In contrast to elastic strain, plastic strain signifies an irreversible process resulting from the movement of dislocation facilitated by the stress at the yield point \citep{Kolsky1949AnLoading}. The yield point is effectively captured by the von Mises yield \replaced{function}{criterion}\chglabel{2.4ba}, represented by \( \varphi \). Specifically, in the context of kinematic hardening, often associated with the Bauschinger effect in metals—the yield \replaced{function}{criterion}\chglabel{2.4bb} is expressed as:

\begin{equation}
\varphi(\boldsymbol{\sigma},\boldsymbol{\alpha}) := \frac{1}{2}\text{dev}(\boldsymbol{\sigma} - \boldsymbol{\alpha}):\text{dev}(\boldsymbol{\sigma} - \boldsymbol{\alpha}) - \frac{1}{3}\sigma_y^2.
\end{equation}
\\
\noindent
Here, \( \sigma_y \) represents the current yield stress, \( \text{dev}(\boldsymbol{\sigma} - \boldsymbol{\alpha}) \) denotes the deviatoric component of the difference between \( \boldsymbol{\sigma} \) and the back stress tensor \( \boldsymbol{\alpha} \).

\begin{figure}[H]
\centering
    \centering
    \includegraphics[width=0.8\textwidth]{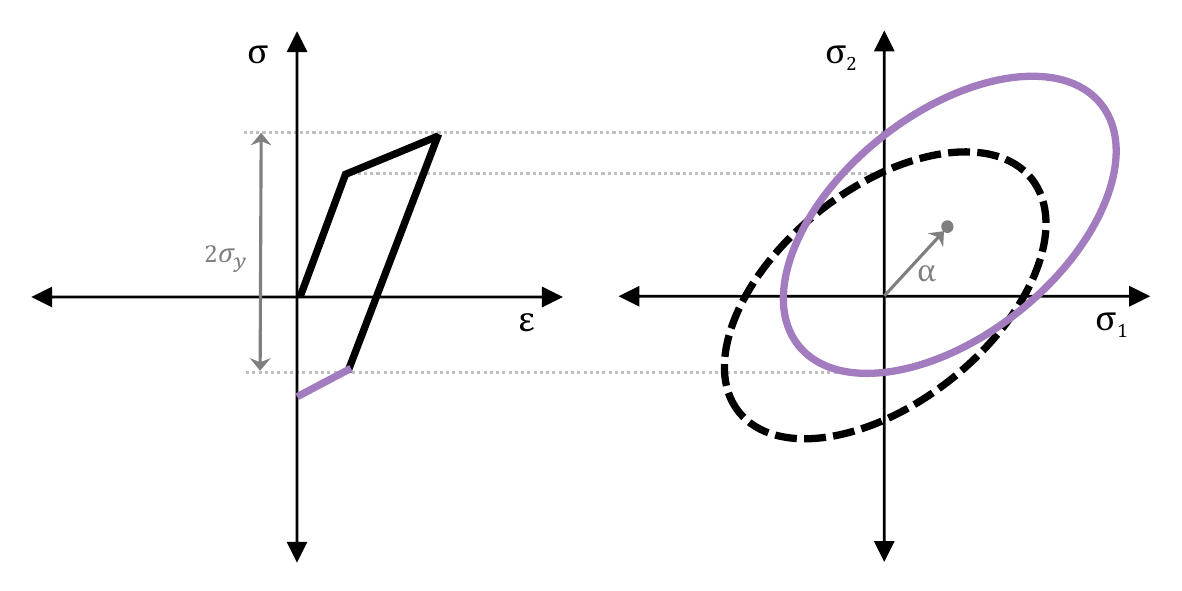}
    \captionsetup{justification=centering} 
    \caption{Kinematic hardening.}
    \label{fig:kinematic_hardening}
\end{figure}

The shift in the yield surface is governed by the back stress tensor $\alpha$, as depicted in Figure \ref{fig:kinematic_hardening}. 

\noindent
The irreversible nature of plastic
flow is captured by means of the Kuhn–Tucker loading and unloading conditions:

\begin{equation}
    \lambda \geq 0, \quad \varphi(\boldsymbol{\sigma}, \boldsymbol{\alpha}) \leq 0, \quad \lambda \varphi(\boldsymbol{\sigma}, \boldsymbol{\alpha}) = 0.
    \label{eq:Kuhn–Tucker}
\end{equation}
\noindent
Here, \( \lambda \) is a non-negative scalar representing the plastic multiplier. The plastic behavior, i.e. the rate of the back-stress tensor 

$\dot{\boldsymbol{\alpha}}$ is characterized by the evolution law:
\begin{equation}
\dot{\boldsymbol{\alpha}}:=H\dot{\boldsymbol{\varepsilon}}_{pl}
\end{equation}
\noindent
Here, \( H \) represents the kinematic hardening modulus and \( \dot{\boldsymbol{\varepsilon}}_{pl} \) the plastic strain rate. In the case of plastic flow, the plastic strain rate tensor can be expressed as:

\begin{equation}
\dot{\boldsymbol{\varepsilon}}_{pl} := \lambda \frac{\partial \varphi}{\partial\boldsymbol{\sigma}}. 
\label{eq:plas_strain_rate}
\end{equation}
\\
\noindent
In the context of incremental plasticity, which calculates the response in discrete steps, the parameter $\lambda$ is fine-tuned at each interval to meet the criteria outlined in \eqref{eq:Kuhn–Tucker}.

In \replaced{summary}{the strong formulation}\chglabel{2.4ca}, the task involves solving for the displacement field \( \boldsymbol{u}: [0, T) \times \Omega \to \mathbb{R}^d \), the tensor of plastic strain expressed as \( \boldsymbol{\varepsilon}_{pl}: [0, T) \times \Omega \to \mathbb{R}^{d \times d}_{\text{sym}} \), and the back stress tensor \( \boldsymbol{\alpha}: [0, T) \times \Omega \to \mathbb{R}^{d \times d}_{\text{sym}} \) such that the following equations hold for all \( t \in [0, T) \)::

\begin{equation}
\left\{
\begin{aligned}
&\nabla \cdot \boldsymbol{\sigma} = 0 \quad &\text{in} \ \Omega, \\
&\boldsymbol{\sigma}  = \mathbb{C}: (\boldsymbol{\varepsilon}  - \boldsymbol{\varepsilon}_{pl}) \quad &\text{in} \ \Omega, \\
&\varphi\left(\boldsymbol{\sigma}, \boldsymbol{\alpha} \right) \leq 0, \quad &\text{in} \ \Omega, \\
&\dot{\boldsymbol{\varepsilon}}_{p}  =  \lambda \frac{\partial \varphi}{\partial \boldsymbol{\sigma}}, \quad &\text{if} \ \varphi = 0 \ \text{and} \ \lambda \geq 0, \text{ in } \Omega, \\
&\lambda \varphi = 0, \quad &\text{in} \ \Omega, \\
&\dot{\boldsymbol{\alpha}} = H\dot{\boldsymbol{\varepsilon}}_{pl}, \quad &\text{in} \ \Omega, \\
&\boldsymbol{u} = \boldsymbol{0} \quad &\text{on} \ \partial \Omega_D, \\
&\boldsymbol{\sigma} \cdot \boldsymbol{n} = \boldsymbol{g}_N \quad &\text{on} \ \partial \Omega_N.
\end{aligned}
\right.
\label{eq:PDEs}
\end{equation}
\\
\noindent
Given the complexity of these equations which represent a rate-independent plasticity with kinematic hardening, a numerical solution via the finite element method (FEM) is often employed. In our case, the PyFEM 
library \citep{Remmers2024PyFEM} serves as the computational tool for iterative solutions. This Python-based finite element code complements the principles and methodologies presented in the book by \citet{deBorst2012Non-LinearEdition}.

\subsection{Parameters and Conditions}
In the remainder of this paper, we explore two scenarios with \(d=2\), under plane stress conditions, where \(\sigma_{zz} = \sigma_{xz} = \sigma_{yz} = 0\). \(\sigma_{zz}\) is the stress normal to the plane, and \(\sigma_{xz}\), \(\sigma_{yz}\) are shear stresses. The FEM simulations employ a nonlinear solver with small strain elements. Material properties follow Table \ref{tab:mechanical_properties}, aligning with \citet{Im2021SurrogateDecomposition} for consistency.

\begin{table}[h]
\centering
\caption{Properties of the used elasto-plastic material.}
\label{tab:mechanical_properties}
\begin{tabular}{p{4cm}lp{0.7cm}l}
\toprule
\textbf{Property} & \textbf{Symbol} & \textbf{Value} & \textbf{Unit} \\
\midrule
Young's Modulus & $E$ & 210 & GPa \\
Poisson's ratio & $\nu$ & 0.3 & - \\
Initial yield stress & $\sigma_y$ & 250 & MPa \\
Kinematic Hardening Modulus & $H$ & 21 & GPa \\
\bottomrule
\end{tabular}
\end{table}

\section{Supervised Learning for Reduced Order Modeling}
To construct a surrogate model that closely follow the finite element method solutions, we propose a deep learning-based mapping. This mapping, denoted as \[
(t, \boldsymbol{\mu}) \mapsto \boldsymbol{a}_h(t, \boldsymbol{\mu}),\] correlates each pair of time \( t \) and parameters \(  \boldsymbol{\mu} \in \mathcal{P} \subset \mathbb{R}^{n_{\mu}} \) with the corresponding high-fidelity numerical solution \replaced{\(\boldsymbol{a}_h(t, \boldsymbol{\mu}) \in \mathbb{R}^{N_h} \)}{\(\boldsymbol{u}_h(t, \boldsymbol{\mu}) \in \mathbb{R}^{N_h} \)}\chglabel{2.4da}, which represents a field variable. We then use this data to approximate the mapping through a data-driven approach.

\subsection{Data}
The solutions are obtained through finite element analysis of the PDEs using space discretization. The dimension of the full order model (FOM) $N_h$ is related to the nodal points and the dimension $d$, where $h > 0$ denotes the discretization parameter controlling the complexity and accuracy of the approximation. The parameter \(\boldsymbol{\mu}\) is generally representative of the physical or geometrical attributes relevant to the problem. In the context of the use cases that will be presented, \(\boldsymbol{\mu}\) includes variations in the applied external force $\boldsymbol{g}_N$.

The goal is to construct an accurate mapping for the entire solution set
\[
\mathcal{S}_h = \{\boldsymbol{a}_h(t; \boldsymbol{\mu}) \mid t \in [0, T) \text{ and } \boldsymbol{\mu} \in \mathcal{P} \subset \mathbb{R}^{n_{\mu}}\} \subset \mathbb{R}^{N_h},
\]
where \((t; \boldsymbol{\mu})\) varies in \([0, T) \times \mathcal{P}\), thereby generating a solution manifold. The solution manifold has to embody the system's history, capturing the evolving states and their dependencies on past conditions and parameters. This means that to fully characterize the dynamics and dependencies of the evolving states within the solution manifold, the dimensionality notation \(n_{\mu} + 1\) is often employed. Here, \(n_{\mu}\) represents the number of parameters \(\boldsymbol{\mu}\) influencing the system, and the "+1" signifies the incorporation of the temporal dimension. This temporal component is crucial for understanding how the system's states progress over time in response to changing parameters and initial conditions. This intrinsic coordinate will be incorporated within the neural network.

\subsubsection{Dimensionality Reduction}
In employing a projection-based reduced order model (ROM) to condense the space \(\mathcal{S}_h \subset \mathbb{R}^{N_h}\), the goal is to identify a lower-dimensional subspace. Within this context, the linear ROM approximate solution \(\tilde{\boldsymbol{a}}_h(t; \boldsymbol{\mu})\) tries to approximate \(\boldsymbol{a}_h(t; \boldsymbol{\mu})\) as:
\begin{equation}
\tilde{\boldsymbol{a}}_h(t; \boldsymbol{\mu}) = \boldsymbol{\Phi}\boldsymbol{a}_n(t; \boldsymbol{\mu}),
\label{eq:linear_ROM}
\end{equation}
with the modes \(\boldsymbol{\Phi} \in \mathbb{R}^{N_h \times n}\) serving as the transformation matrix by projecting the full space into a n-dimensional subspace $\mathcal{\tilde{S}}_n \subset \mathbb{R}^{N_h}$. If $\boldsymbol{\Phi}$ has full rank, this subspace has a reduced rank \(n\), satisfying \(n \ll N_h\). The vector of eigenvalues is denoted by \replaced{\(\boldsymbol{a}_n(t; \boldsymbol{\mu}) \in \mathbb{R}^n\)}{\(\boldsymbol{u}_n(t; \boldsymbol{\mu}) \in \mathbb{R}^n\)}\chglabel{2.4db}, which consists of the intrinsic degrees of freedom of the ROM. Consequently, the mapping \(\tilde{\boldsymbol{a}}_h : [0, T) \times \mathcal{P} \rightarrow \mathcal{\tilde{S}}_n\) is defined, where \(\mathcal{\tilde{S}}_n\) symbolizes the reduced linear trial manifold, characterized as
\begin{equation}
\tilde{S}_n = \text{Im}(\boldsymbol{\Phi}).
\end{equation}

\noindent
In the context of POD-based ROMs, this space is spanned by the first $n$ singular vectors from the snapshot matrix
\begin{equation}
   \boldsymbol{S} := \left[ \boldsymbol{a}_h(t^1, \boldsymbol{\mu}_1) | \ldots | \boldsymbol{a}_h(t^{N_t}, \boldsymbol{\mu}_1)| \ldots |  \boldsymbol{a}_h(t^1, \boldsymbol{\mu}_{N_{\text{train}}})  | \ldots |  \boldsymbol{a}_h(t^{N_t}, \boldsymbol{\mu}_{N_{\text{train}}}) \right] \in \mathbb{R}^{N_h \times N_s}.
\end{equation}
This matrix compiles a sequence of snapshot arrays derived from the Full Order Model (FOM) solutions at distinct temporal intervals, denoted by \(\{t^k\}_{k=0}^{N_t}\), with each interval \(t^k\) specified as \(k\Delta t\). The \replaced{load step}{timestep}\chglabel{2.4ga} is defined by \(\Delta t = \frac{T}{N_t}\), and \(N_s = N_{set} \times N_t\), with the parameter values \(\boldsymbol{\mu} \in \mathcal{P}\), where \(N_{set}\) signifies the number of samples within a specific set.

For the computation of the POD modes and eigenvalues, analogous to the procedure in PCA, we first centralize the matrix \(\boldsymbol{S}\) by calculating its column-wise mean vector \(\bm{\bar{s}} \in \mathbb{R}^{N_h}\). Then, to form \(\bm{\bar{S}} \in \mathbb{R}^{N_h \times N_s}\), we replicate \(\bm{\bar{s}}\) across \(N_s\) columns, thereby constructing a matrix where each column is the mean vector \(\bm{\bar{s}}\). Subsequently, this mean vector is deducted from \(\boldsymbol{S}\) to derive the deviation matrix \(\boldsymbol{S}' = \boldsymbol{S} - \boldsymbol{\bar{S}}\).\added{Unlike PCA, in which the covariance matrix is calculated, we apply SVD directly on the centered data \(\boldsymbol{S}'\) to capture the dominant modes}\chglabel{2.4ea}.
The subsequent step involves decomposing the centered data via Singular Value Decomposition (SVD), which can be expressed as:
\begin{equation}
\bm{S}' = \bm{\Phi} \bm{\Pi} \bm{\Psi}^\top.
\end{equation}
In this decomposition, the orthonormal basis for the column space of $\bm{S}'$ is formed by the columns of $\bm{\Phi}= [\boldsymbol{\phi}_1|\ldots|\boldsymbol{\phi}_{N_h}] \in \mathbb{R} ^{N_h \times N_h}$, which are key in encoding spatial patterns. Similarly, the columns of $\bm{\Psi}=[\boldsymbol{\psi}_1|\ldots|\boldsymbol{\psi}_{N_s}] \in \mathbb{R}^{N_s \times N_s}$, establish an orthonormal basis for the row space of $\bm{S}'$ representing temporal patterns. Additionally, $\bm{\Pi}=\text{diag}(\pi_1,\ldots,\pi_r) \in \mathbb{R}^{N_h \times N_s}$ with $\pi_1 \geq \pi_2 \geq \ldots \geq \pi_r$, and $r \leq \text{min}(N_h,N_s)$. 

The orthonormal characteristic of $\bm{\Phi}$ ensures that the matrices' inverses are identical to their transposes, as indicated in \citet{Brunton2019Data-DrivenEngineering}, facilitating the expression of \replaced{$\bm{A}$}{$\bm{U}$}\chglabel{2.4dc}$= [\bm{\phi}_1^\top \bm{S}' | \ldots | \bm{\phi}_r^\top \bm{S}'] \in \mathbb{R}^{r \times N_s}$. This orthonormality of $\bm{A}$'s columns assures the most accurate reconstruction of the snapshot data within any $r$-dimensional subspace \citep{Fresca2021APDEs}.

To condense the subspace from $r$ dimensions to $n$ dimensions, where $n \ll r$, the approach involves selecting the $n$ leading modes, from $\bm{\Phi}$, resulting in $\bm{\Phi}_n = [\bm{\phi}_1 | \cdots | \bm{\phi}_n] \in \mathbb{R}^{N_h \times n}$. Similarly, the selection of the $n$ eigenvalues, is achieved with $\bm{A}_n = [\bm{a}_1 | \cdots | \bm{a}_n] = [\bm{\phi}_1^\top \bm{S}' | \cdots | \bm{\phi}_n^\top \bm{S}'] \in \mathbb{R}^{n \times N_s}$. Following the linear dimensionality reduction, the reconstruction of the snapshot matrix yields:
\begin{equation}
\tilde{\bm{A}}_h = \tilde{\bm{S}} = \bm{\Phi}_n \bm{A}_n + \bar{\bm{S}}. 
\end{equation}
Determining the dominant vectors $n$ is done with use of a criteria. We use the a percentage-based Root Mean Square Error $e_{\text{RMSE}\%}$, which involves incrementally increasing $n$ until the $e_{\text{RMSE}\%}$ between the FOM data $\bm{S}$ and the ROM data $\tilde{\bm{S}}$ falls below a specified threshold:
\begin{equation}
e_{\text{RMSE}\%} = \frac{1}{N_{s}} \sum^{N_{s}}_{j=1} \left(\frac{1}{\underset{i = 1, \ldots, N_h}{\max}|\bm{S}_{ij}|} \sqrt{\frac{\sum^{N_h}_{i=1}(\bm{S}_{ij}  - \tilde{\bm{S}}_{ij})^2}{N_h}}\right) \times 100\%.
\end{equation}
Here, $i$ indexes the rows, while 
$j$ indexes the columns.

\subsubsection{Data Collection \& Preprocessing}
To create a FOM dataset for multiple field variables as listed in Table \ref{tab:field_variables}, the parameters are sampled randomly from the subspace \(\mathcal{P} \subset \mathbb{R}^{n_\mu}\), creating a parameter matrix \(\bm{X}=[\bm{x}_1^\top|\ldots|\bm{x}_{N_s}^\top]  \in \mathbb{R}^{n_\mu \times N_s} \). Note that the plastic strain tensor has \replaced{6 components, but 2 }{6 dimensions, but 3}\chglabel{2.4fa} components are zero due to PyFEM's conversion from a 3D to a 2D situation, resulting in a sparse representation.

\begin{table}[h]
\centering
\caption{Field variables.}
\label{tab:field_variables}
\begin{tabular}{p{3.4cm}lp{2.3cm}l} 
\toprule
\textbf{Variable} & \textbf{Symbol} & \textbf{Dimensions ($d$)} & \textbf{Unit} \\
\midrule
Displacement & $\bm{u}$ & 2 & mm \\
Von Mises Stress & $\sigma_{\text{vm}}$ & 1 & MPa \\
Equivalent plastic strain & $\bar{\varepsilon}_{\text{pl}}$ & 1 & - \\
Plastic strain tensor & $\bm{\varepsilon}_{\text{pl}}$ & 6 & - \\
\bottomrule
\end{tabular}
\end{table}
\noindent

This process results in multiple FOM solutions for various field variables, which are then reduced in size by setting \(e_{\text{RMSE}\%} < 1\%\), yielding \(\bm{A}_n^i\) for each \(i\)-th field variable. 
These ROMs, \(\bm{A}_n^i\), are assembled into a unified eigenvalue tensor \(\bm{Y} = [\bm{y}_1^\top|\ldots|\bm{y}_{N_s}^\top] = [\bm{A}_1| \ldots | \bm{A}_{N_n}]\)  \(\in \mathbb{R}^{N_n \times N_s}\). 
Here, the eigenvalues are stacked such that \( N_n = \sum_{i=1}^{N_f} n^i \), where \( N_f \) is the number of fields and \( n^i \) is the number of eigenvalues.
For each use case, we have a dataset of 300 samples, with a time span defined by \(N_t=10\). These samples are divided such that 80\% (\(N_{train}=240\)) are dedicated to the training set and 20\% (\(N_{val}=60\)) to the validation set. A test set of \(N_{test} = 5\) samples, chosen to reflect realistic scenarios, is illustrated in detail in Table \ref{tab:test_set} and Figure \ref{fig:force_trend}
\begin{table}[h]
\centering
\begin{minipage}{0.35\textwidth}
    \centering
    \caption{Naming of the cases.}
    \label{tab:test_set}
    \begin{tabular}{p{0.15cm}p{4cm}}
    \toprule
     & \textbf{Loading Case} \\
    \midrule
    1. & Ramp Up \\
    2. & Bidirectional Cyclic Loading \\
    3. & Unidirectional Cyclic Loading \\
    4. & Constant Load \\
    5. & Impulse Load \\
    \bottomrule
    \end{tabular}
\end{minipage}
\hfill
\begin{minipage}{0.6\textwidth}
    \centering
    \includegraphics[width=0.7\textwidth]{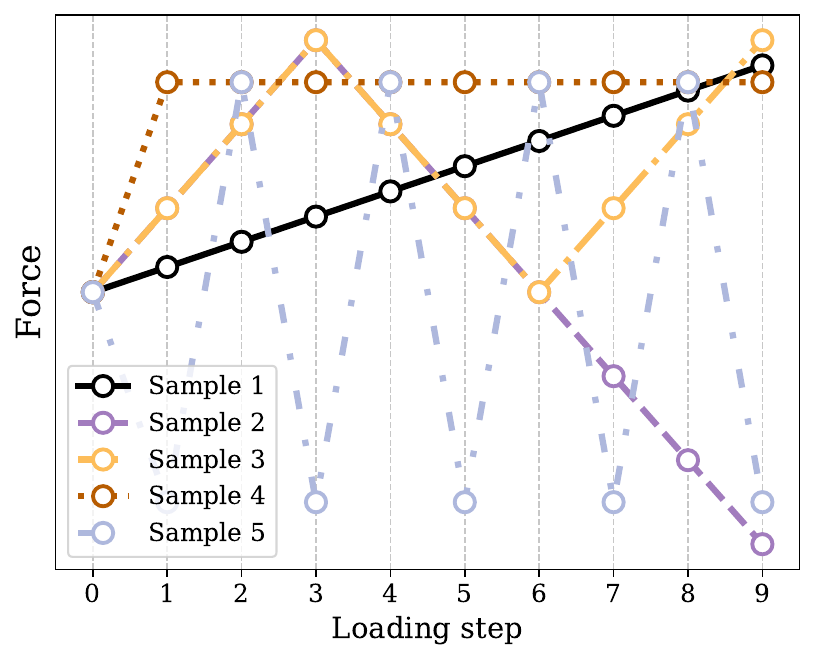}
    \captionsetup{type=figure}
    \captionsetup{justification=centering} 
    \caption{Trends of the force at each loading step.}
    \label{fig:force_trend}
\end{minipage}
\end{table}

\noindent
Note that the Y-axis ticks are omitted to emphasize the trend, as the magnitudes are dependent on the problem context. Thereafter it is important to mention that the initial condition for each sample sets the force magnitude to zero, resulting in \(\bm{a}_n(t^0;\bm{\mu})=\bm{0}\).

The parameter matrices for every dataset $\bm{X}^\text{set}$ are subjected to min-max normalization, which is chosen for its suitability with uniformly distributed data. The equation used rescale the features of the normalized input $\mathbf{X}$ is given by:
\begin{equation}
\begin{aligned}
\mathbf{X}_{ij}^\text{set} &\mapsto \frac{\mathbf{X}_{ij}^\text{set} - \underset{j=1,...,N_s}{\max} (\mathbf{X}_{ij}^{\text{train}})}{\underset{j=1,...,N_s}{\max} (\mathbf{X}_{ij}^{\text{train}})
 - \underset{j=1,...,N_s}{\min} (\mathbf{X}_{ij}^{\text{train}})}, & i &= 1, \ldots, n_\mu, & j &= 1, \ldots, N_s.
\end{aligned}
\end{equation}
\noindent
Every feature within the composite eigenvalue tensors for all the datasets are standardized according to Z-score standardization:
\begin{equation}
\begin{aligned}
\mathbf{Y}_{ijk}^\text{set} &\mapsto \frac{\bm{Y}_{ijk}^\text{set} - \bar{\bm{Y}}^{\text{train}}_{ij}}{\overline{\bm{Y}}_{ij}^\text{train}}, & i &= 1, \ldots, N_v, & j &= 1, \ldots, n & k &= 1, \ldots, N_s.
\end{aligned}
\end{equation}

\noindent
In this equation, $\bar{\bm{Y}}_{ij}^\text{train}$ signifies the mean matrix, and $\overline{\bm{Y}}_{ij}^\text{train}$ signifies the standard deviation of $\bm{Y}_{ij}^\text{train}$. 

\subsection{Neural Networks}
With use of the data obtained through the POD method, we establish a mapping represented by a neural network $\bm{f}^\text{NN}$. This network comprises \(L\) layers; each layer is characterized by its own set of biases \(\bm{b} \in \mathbb{R}^{M}\) and is linked with the subsequent layer through a matrix of weights \(\bm{W} \in \mathbb{R}^{M \times D}\), where \(D\) represents the number of input features, and \(M\) denotes the number of hidden units. The expression for each layer is given by:
\begin{equation}
\bm{h}_1(\bm{x}) = \bm{f}_1(\bm{W}_0\bm{x} + \bm{b}_0) \quad \bm{h}_{l+1}(\bm{h}_l) = \bm{f}_{l+1}(\bm{W}_l\bm{h}_l + \bm{b}_l).
\end{equation}

\noindent
In these equations, $\bm{f}_1$ denotes the first activation function, and $\bm{x}$ represents the input to the first layer $\bm{h}_1$ with $D=N_\mu$. For the next hidden state $\bm{h}_{l+1}$ the input of its previous hidden state $\bm{h}_{l}$ is used at the $l$-th layer.

Given the neural network's parameters $\bm{\theta}$, encompassing both weights and biases such that $\boldsymbol{\theta} = \{\boldsymbol{W}_l, \boldsymbol{b}_l\}_{l=0}^{L-1}$, the network's final output is obtained by sequentially applying layer-wise transformations:
\begin{equation}
\boldsymbol{f}^{\text{NN}}(\boldsymbol{x}) = \boldsymbol{h}_L \circ \boldsymbol{h}_{L-1}  \ldots \circ \boldsymbol{h}_2 \circ \boldsymbol{h}_1(\boldsymbol{x}).
\end{equation}
\noindent
In this context, the term \(\boldsymbol{h}_{L} = \bm{f}_L(\bm{W}_{L-1}\bm{h}_{L-1} + \bm{b}_{L-1})\) signifies the output layer, with \(M = n\) ensuring the desired dimensionality of the output. Additionally, the notation \(\circ\) denotes the composition of functions.

Given a \replaced{load step}{timestep}\chglabel{2.4gb} $t$ and its current parameters $\boldsymbol{\mu}$, we define a map which the neural network has to approximate: 
\[(t, \boldsymbol{\mu}) \mapsto \boldsymbol{a}_n(t; \boldsymbol{\mu}).\] 
By optimizing $\bm{\theta}$, the aim is to approximate  the mapping. This is done by minimizing the defined loss function:
\begin{equation}
\mathcal{J}(\bm{\theta}) = \frac{1}{N_s} \sum_{i=1}^{N_{\text{train}}} \sum_{k=1}^{N_t} \mathcal{L}(t^k, \bm{\mu}_i; \bm{\theta}) \rightarrow \min_{\bm{\theta}}
\end{equation}
where
\begin{equation}
\mathcal{L}(t^k, \bm{\mu}_i; \bm{\theta}) = 
\|\bm{a}_n(t^k, \bm{\mu}_i) - \boldsymbol{f}^\text{NN}(t^k;\boldsymbol{\mu_i},\bm{\theta})\|^2.
\end{equation}

From the minimization process, we obtain an optimal \(\theta^*\).
The NN approximation is then defined as:
\begin{equation}
(t, \boldsymbol{\mu})\mapsto
\tilde{\boldsymbol{a}}_n(t; \boldsymbol{\mu}):= \boldsymbol{f}^\text{NN}(t;\boldsymbol{\mu},\bm{\theta}).
\end{equation}
\subsubsection{Long Short-Term Memory}
Modeling elastoplastic materials requires incorporating a history component, which standard NNs lack. This means that the mapping has to include some parameter accounting for its intrinsic dimension. To resolve this, a Recurrent Neural Network (RNN) is employed for its recursive data processing capability. In addition to receiving the input of the previous layer $\bm{x}$, RNN neurons also receive the output of the current layer at time $t^{k-1}$ such that the expression of each layer is given by:
\begin{align}
\bm{h}^0 &= \bm{f}(\bm{W}\bm{x}^0 + \bm{b}), \\
\bm{h}^k &= \bm{f}(\bm{W}\bm{x}^k + \bm{R}\bm{h}^{k-1} + \bm{b}),
\end{align}
where $\bm{R} \in \mathbb{R}^{M \times M}$ is the hidden weight matrix accounting for the hidden state at $t^{k-1}$.

\begin{figure}[H]
\centering
\includegraphics[width=0.75\textwidth]{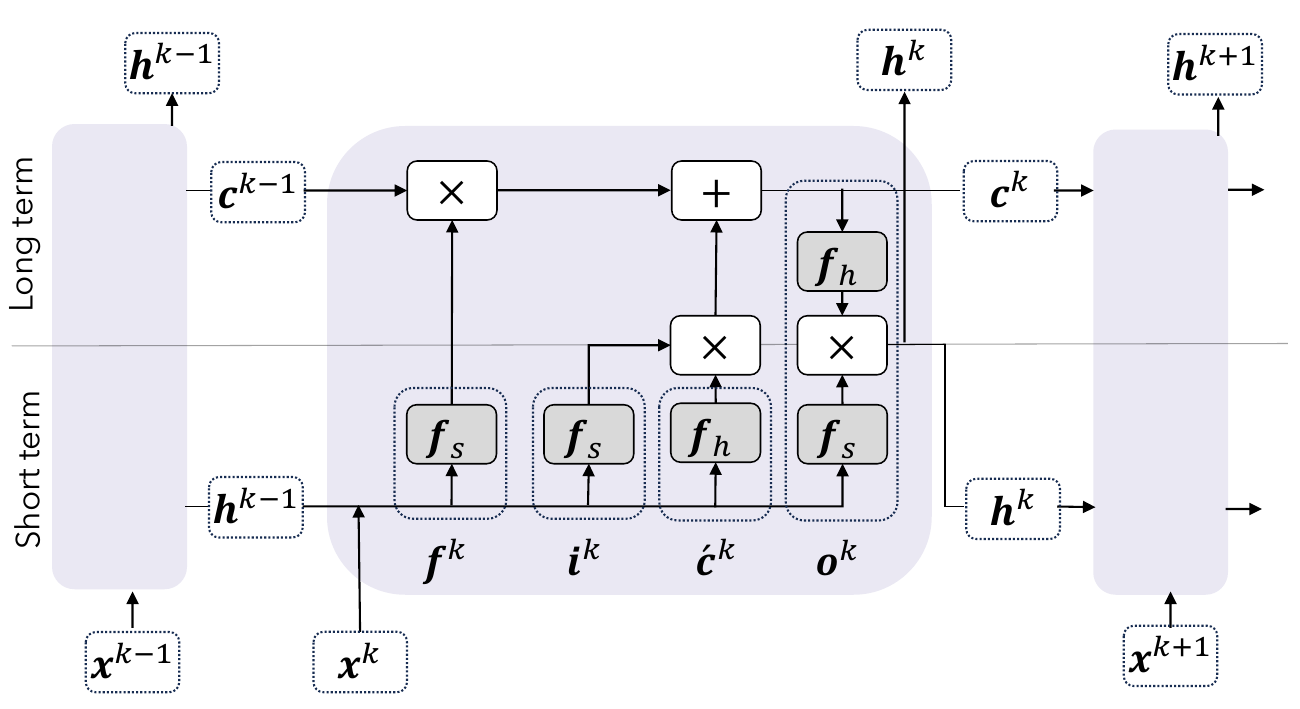}
\captionsetup{justification=centering} 
\caption{LSTM cell highlighting trainable parameters in shaded areas.}
\label{fig:LSTM}
\end{figure}

Unfortunately, simple RNNs suffer from the vanishing gradient problem \citep{Ribeiro2020BeyondSmoothness}, where the internal state diminishes as the sequence length increases, making it challenging to compute gradients and train the network. 
\noindent
To address this, Long Short-Term Memory (LSTM) neural networks have been devised, integrating two distinct streams \citep{Hochreiter1997LongMemory}. On the one hand, the intrinsic dimension has to be captured by the cell state $\bm{c}^k$, responsible for long-term memory, while short-term memory incorporates the hidden state $\bm{h}_t$ as illustrated in \ref{fig:LSTM}.

\vspace{-0.2cm}
\noindent
The input gate $\bm{i}^k$, the cell input $\Acute{\bm{c}}^k$, output gate $\bm{o}^k$, and forget gate $\bm{f}^k$ are responsible for updating this history component, as described by the following equations: 
\begin{equation}
\begin{aligned}
\bm{f}^k & =\bm{f}_s\left(\bm{W}_{f} \bm{x}^k+\bm{R}_{f} \bm{h}^{k-1}+\bm{b}_f\right), \\
\bm{i}^k & =\bm{f}_s\left(\bm{W}_{i} \bm{x}^k+\bm{R}_{i} \bm{h}^{k-1}+\bm{b}_i\right), \\
\bm{o}^k & =\bm{f}_s\left(\bm{W}_{o} \bm{x}^k+\bm{R}_{o} \bm{h}^{k-1}+\bm{b}_o\right), \\
\Acute{\bm{c}}^k & =\bm{f}_h\left(\bm{W}_{c} \bm{x}^k+\bm{R}_{c} \bm{h}^{k-1}+\bm{b}_c\right), \\
\bm{c}^k & =\bm{f}^k \odot \bm{c}^{k-1}+\bm{i}^k \odot \Acute{\bm{c}}^k, \\
\bm{h}^k & =\bm{o}^k \odot \bm{f}_h\left(\bm{c}^k\right)
\end{aligned}
\end{equation}
In the presented equations, the sigmoid and hyperbolic tangent functions are represented by \(\bm{f}_s\) and \(\bm{f}_h\), respectively. The sigmoid function, which outputs values between 0 and 1 manages the flow of information, which is vital for sustaining memory and ensuring gradient stability. Conversely, the hyperbolic tangent function, yielding values from -1 to 1, controls the dynamics within cell and hidden states, facilitating the modeling of long-term dependencies. Furthermore, the symbol \(\odot\) is used to denote element-wise multiplication.

\subsubsection{Multi-Task Learning}
In addtion to the approach used by \citet{Im2021SurrogateDecomposition}, our method aims to comprehensively understand the problem's dynamics by utilizing multiple fields. Thus, the output is arranged as described earlier by the composite eigenvalue tensor \( \bm{Y} \). This tensor's components concatenates the eigenvalues at a particular \replaced{load step}{time step}\chglabel{2.4gc} such that \( \bm{y} = \begin{bmatrix} \bm{y}_n^{1} & \cdots & \bm{y}_n^{N_f} \end{bmatrix}
 \in \mathbb{R}^{N_n} \). These eigenvalues are extracted from the FOM given by \( \bm{a}_h^i : [0, T) \times \mathcal{P} \rightarrow \mathbb{R}^{N_h^i} \). 

\begin{figure}[H]
\centering
\includegraphics[width=0.70\textwidth]{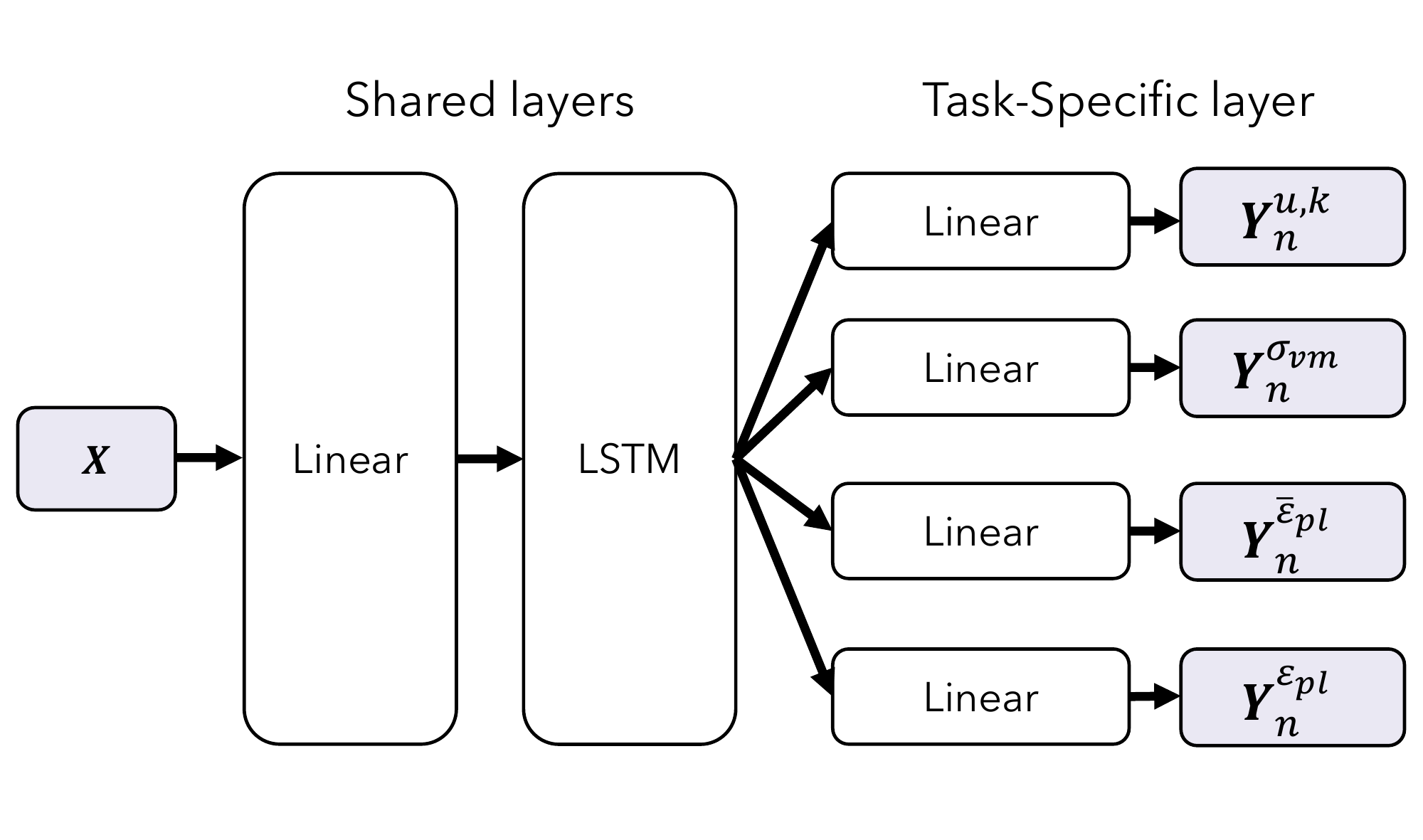}
\captionsetup{justification=centering} 
\caption{Architecture of the neural network.}
\label{fig:MTL_arch}
\end{figure}

Given the significant correlation among these variable fields, using a multi-task network is advantageous. This approach enables learning a common representation of the problem through its shared layers, rather than relying on a single-task neural network. The design of this network, as adapted for our specific application, is depicted in Figure \ref{fig:MTL_arch}.
\noindent
This incentivizes the model to discover a comprehensive latent representation that encompasses the common underlying factors and interrelationships among various outputs.
Concurrently, the distinct task-specific output layers maintain the adaptability necessary for specialization in each field variable.

This strategy, widely recognized as Multi-Task Learning (MTL), falls within the domain of ML subfields and was initially introduced by \citet{Caruana1997multitaskLearning}. Functioning as a type of transfer learning, MTL enables a model to simultaneously grasp multiple interconnected tasks, providing several advantages such as improved data efficiency and accelerated learning aided by auxiliary information \citep{Vandenhende2022Multi-TaskSurvey}. Subsequently \citet{Baxter1997ASampling}, has shown that the risk of overfitting training data is considerably lower compared to its single-task equivalent. This aligns with the intuitive idea that, as the model learns more tasks concurrently, it must discover a representation that accommodates them all, thereby reducing the likelihood of overfitting the original task.

\subsubsection{Hyperparameter Tuning}
Optimization of NN's hyperparameters, such as the LSTM layer's hidden state size \( H \), is crucial for the model's performance and is achieved through systematic tuning \citep{Yu2020Hyper-ParameterApplications}. Utilizing Optuna's Tree-structured Parzen Estimator (TPE) for Bayesian optimization allows efficient navigation through complex hyperparameter spaces by assessing the probability of hyperparameter sets \( \boldsymbol{\beta} \) against objective function values \( \tau \) \citep{Akiba2019Optuna:Framework, Watanabe2023Tree-StructuredPerformance}.

TPE uses probabilistic models \( l(\boldsymbol{\beta}) \) and \( g(\boldsymbol{\beta}) \) for kernel density estimation to identify superior hyperparameter configurations. An acquisition function, typically the expected improvement ratio \( \frac{l(\boldsymbol{\beta})}{g(\boldsymbol{\beta})} \), drives the selection process to balance exploration and exploitation effectively.
\begin{equation}
\boldsymbol{\beta}_{\text{opt}} = \arg\max_{\boldsymbol{\beta}} \left( \frac{l(\boldsymbol{\beta})}{g(\boldsymbol{\beta})} \right)
\end{equation}
This formula encapsulates TPE's strategy to find optimal hyperparameters that improve the mean RMSE of the validation set, verified over multiple training rounds of 5000 epochs for reliability. Defined ranges for each hyperparameter ensure a structured search, as detailed in Table \ref{tab:Hyperparameters}.

\begin{table}[h]
\centering
\caption{Hyperparameter ranges.}
\label{tab:Hyperparameters}
\begin{tabular}{p{4.5cm}p{3.4cm}} 
\toprule
Hyperparameter& Range \\
\midrule
Batch Size & [1, 128] \\
LSTM Hidden State Size ($H_\text{LSTM}$) & [$N_n$, 100] \\
Learning Rate & [\num{1e-3}, \num{1e-1}] \\
Number of LSTM Layers & [1,5] \\
Weight Decay & [\num{1e-7}, \num{1e-4}] \\
\bottomrule
\end{tabular}
\end{table}
\noindent

\subsubsection{Evaluation}
During the evaluation phase, multiple independently trained neural networks are evaluated, each characterized by a unique set of parameters represented as \(\bm{\Theta} = [\bm{\theta}_1, \ldots, \bm{\theta}_{N_e}]\), where \(N_e=5\) signifies the total number of networks within the ensemble. \added{These networks are trained on different batches, with samples randomly selected from the entire training set. In the evaluation of networks with varying training sizes, the training samples differ across networks, which increases the robustness of the results.}\chglabel{1.4a} Variability in network outputs leads to the application of an averaging technique:
\begin{equation}
    \bm{y}_n^i \mapsto \frac{1}{N_e}\sum^{N_e}_{e=1} \bm{y}_n^i(t;\bm{\mu},\bm{\theta}_e ),
\end{equation}
with this averaging the robustness and regularization are enhanced \citep{Li2021NeuralDiversity,Ganaie2022EnsembleReview}.

Performance evaluation employs two primary metrics: the relative Mean Absolute Error (MAE\%) and a weighted variant of the coefficient of determination (\(R^2\)). The MAE\% assesses the relative error between the predicted ROM output, \(\hat{\bm{a}}^i_h = \bm{\Phi}^i_n \bm{y}^i_n + \bar{\bm{s}}^i\), and the actual ROM output, \(\tilde{\bm{a}}^i_h = \bm{\Phi}^i_n \bm{a}^i_n + \bar{\bm{s}}^i\), at each integration point \(h\) for a specific field variable \(i\) during a single \replaced{load step}{timestep}\chglabel{2.4gd} \(k\):
\begin{equation}
\bm{\epsilon}_h^{i,k}(\tilde{\bm{a}}^i_h, \hat{\bm{a}}^i_h) = 
\frac{
\left|\tilde{\bm{a}}^{i,k}_h(\bm{\mu}_{\text{test}}) - \hat{\bm{a}}^{i,k}_h(\bm{\mu}_{\text{test}}) \right|}
{\tilde{a}^i_\text{max}(\bm{\mu}_\text{train})}.
\end{equation}

This illustrates the methodology for quantifying error magnitude at each \replaced{load step}{timestep}\chglabel{2.4e} and integration point. The maximum value of the training set for each variable is denoted by $\tilde{a}^i_\text{max}(\bm{\mu}_\text{train})$. It is critical to address scenarios in the test set wherein the maximum value of a variable remains zero for all \replaced{load steps}{time steps}\chglabel{2.4f} within a sample. \replaced{This situation often occurs when the plastic strain remains zero because the yield stress is not exceeded, which can cause issues during normalization due to a zero range in plastic strain values.}{This situation often occurs in cases like plastic strain not exceeding the yield stress, leading to a division by zero.}\chglabel{1.9a} To mitigate this, the maximum of each element is assigned a predefined non-zero value to ensure numerical stability:
\begin{equation}
\tilde{a}^i_{\text{max}}(\bm{\mu}_\text{train}) = \max_{j \in \{1,...,N_\text{train}\}} \max_{k \in \{1,...,N_t\}} \max_{h \in \{1,...,N_h^i\}} \left|\tilde{\bm{a}}^{i,k}_h(\bm{\mu}_{\text{train},j}) \right|.
\end{equation}
The term $\bm{\epsilon}_h^{i,k}(\tilde{\bm{a}}^i_h, \hat{\bm{a}}^i_h)$ denotes the relative error between the predicted and actual values for a single \replaced{load step}{timestep}\chglabel{2.4ge} and field variable at the integration points. It is essential to note that this error metric can be averaged across all quantities, consistently referred to as MAE\%.

The predictive precision of the eigenvalues compared to the ground truth is assessed using the coefficient of determination ($\text{R}^2$), evaluating the neural network's capacity to explain the variance of the eigenvalues. Due to the higher importance of the first eigenvalue in representing the variable, $R^2$ is weighted accordingly:

\begin{equation}
{\bm{R}_\text{weight}^2}^i = 
\sum_{n=1}^{n^i} 
\frac{\bm{w}_n^i (\bm{\mu}_\text{train})}{N_\text{test}}
\sum_{j=1}^{N_\text{test}}
\left( 1 - \frac{\sum_{k=1}^{N_t} ({\bm{y}}_{n}^{i,k} (\bm{\mu}_\text{test,j}) - \bar{{\bm{a}}}^i_{n}(\bm{\mu}_\text{test,j}))^2}{\sum_{k=1}^{N_t} ({\bm{a}}_{n}^{i,k}(\bm{\mu}_\text{test,j}) - {\bar{\bm{a}}^i_{n}}(\bm{\mu}_\text{test,j}))^2} \right).
\end{equation}
In this equation, $\bar{\bm{a}}^i_n \in \mathbb{R}^n$ represents the mean over $k$ of the true eigenvalues while 
$\bm{y}_n^{i,k}$ signifies the predicted eigenvalue at a particular \replaced{load step}{timestep}\chglabel{2.4gf}. The $R^2$ value gets weighted by $\bm{w}_n^i$ denoted by:
\begin{equation}
\bm{w}_n^i(\bm{\mu}_\text{train})=\frac{{\bm{a}}^i_{n,\text{max}} - {\bm{a}}^i_{n,\text{min}}}{\sum_{n=1}^{n^i} ({\bm{a}}^i_{n,\text{max}} - {\bm{a}}^i_{n,\text{min}})},
\end{equation}
where $\bm{a}^i_{n,\text{max}}$ and $\bm{a}^i_{n,\text{min}}$ denote the maximum and minimum values of all \replaced{load steps}{timestep}\chglabel{2.4gg} within the training set, respectively:
\begin{align}
    \bm{a}^i_{n,\text{max}}(\bm{\mu}_\text{train}) &= \max_{j \in \{1,\ldots,N_\text{train}\}} \max_{k \in \{1,\ldots,N_t\}} \left|\bm{a}^{i,k}_n(\bm{\mu}_{\text{train},j}) \right|, \\
    \bm{a}^i_{n,\text{min}}(\bm{\mu}_\text{train}) &= \min_{j \in \{1,\ldots,N_\text{train}\}} \min_{k \in \{1,\ldots,N_t\}} \left|\bm{a}^{i,k}_n(\bm{\mu}_{\text{train},j}) \right|
\end{align}
Reporting both MAE\% and the Weighted $R^2$ for every field variable on the test sets provides a comprehensive understanding of the model performance. The goal is to achieve low MAE\% and high $R^2$ (towards 1) values across both sets, which indicate accurate predictions and strong generalization.
\section{Results and Discussion}
\label{sec4}
The performance of the proposed multi-task LSTM framework was assessed using two two-dimensional scenarios: a table model and a cantilever beam model. The outcomes for each scenario highlight the framework's capability to precisely predict the field variables in the test set instances.

\subsection{Use Case 1: Table}\label{subsec1}
\begin{figure}[H]
    \begin{minipage}{1\textwidth}
        \centering
        \includegraphics[width=0.6\linewidth]{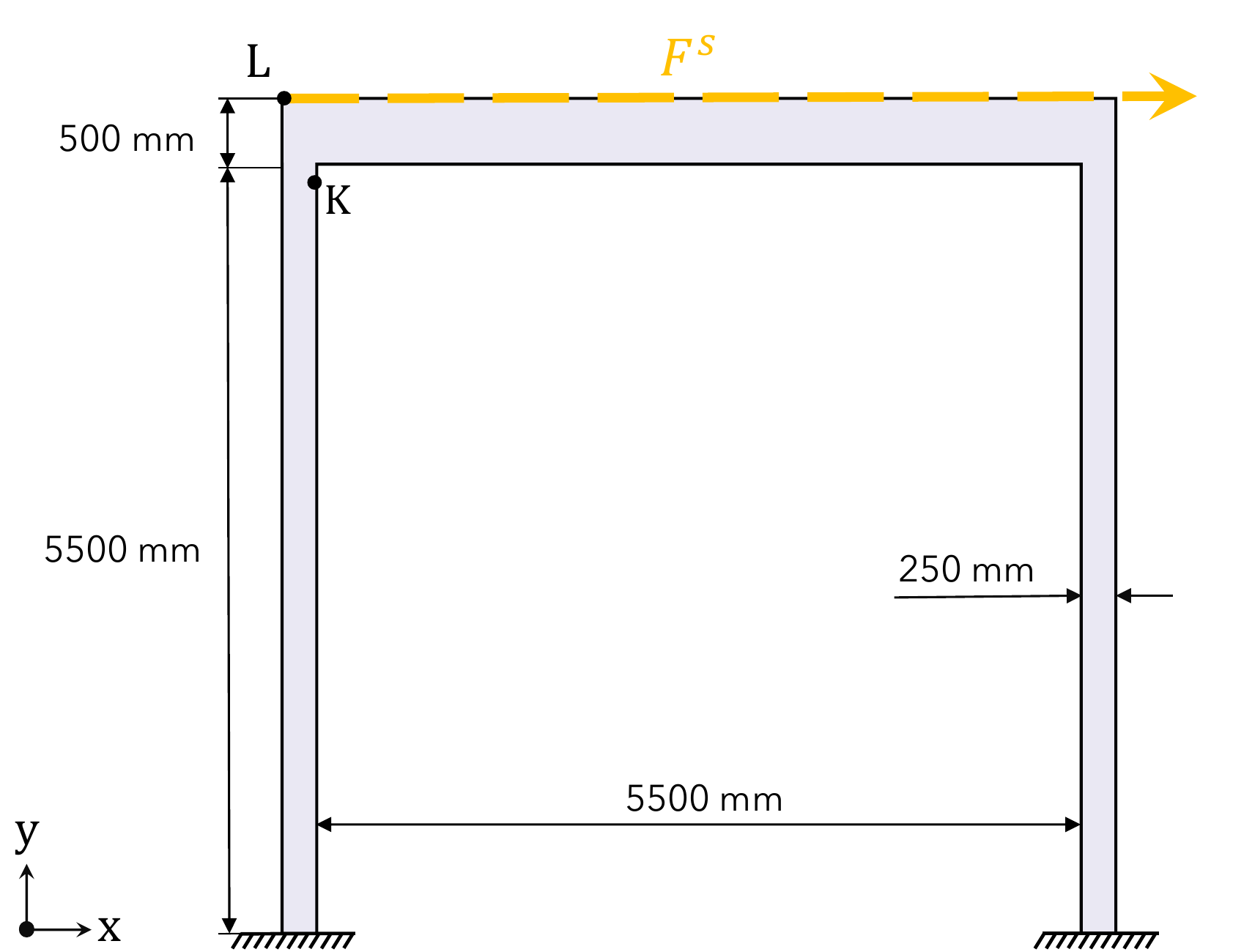}
        \captionsetup{justification=centering} 
        \caption{Table model.}
        \label{fig:table_geom}
    \end{minipage}%
\end{figure}

The proposed framework's effectiveness was initially tested with a 2D table model, replicating the benchmark used by \citet{Im2021SurrogateDecomposition} to demonstrate neural network performance. The FEM model consists of 800 linear 2D quadrilateral elements, with 50 x 5 elements in the tabletop and 5 x 50 elements in each table leg. The geometry and boundary conditions of the table, are illustrated in Figure \ref{fig:table_geom}. 
\noindent
The training and validation datasets are generated by applying a random shear force $F^s$ to the top surface of the table for $N_t=10$ \replaced{load steps}{timesteps}\chglabel{2.4gh}, with $N_\mu=1$ within the range of \replaced{$\mathcal{P}=[-3.5,3.5] \text{ N/mm}$}{$\mathcal{P}=[3.5,3.5] \text{ N/mm}$}\chglabel{2.4ha} . This implies that the variable boundary condition consists exclusively of a changing shear force in the specified direction.
\subsubsection{Data Reduction}
\begin{figure}[H]
    \centering
    \begin{minipage}{0.5\textwidth}
        \centering
        \includegraphics[width=\linewidth]{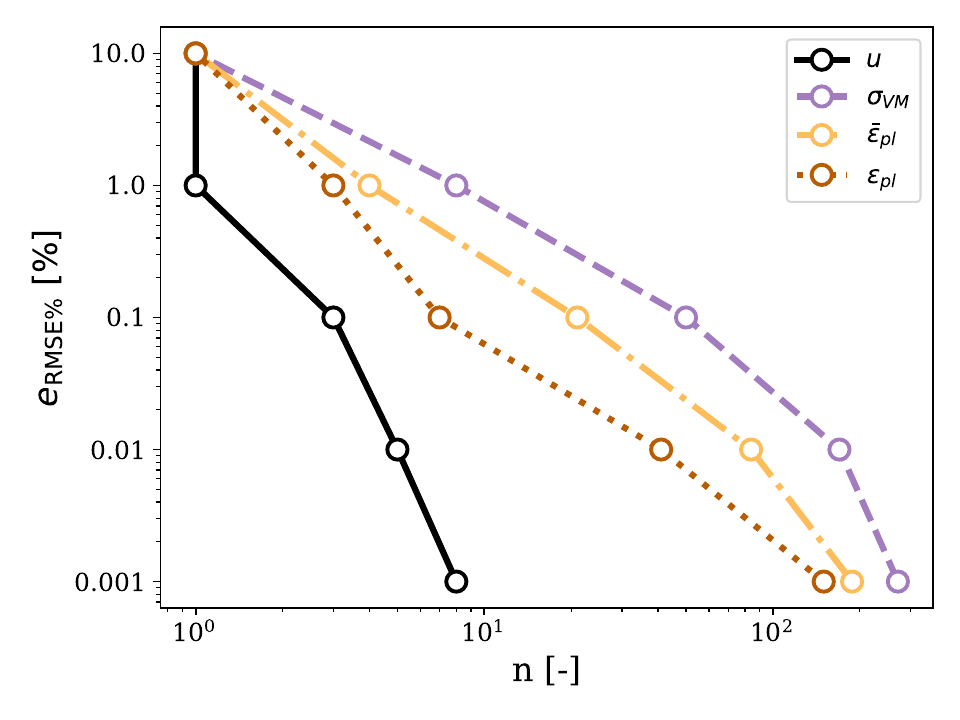}
        \caption{POD accuracy vs efficiency.}
        \label{fig:POD_table}
    \end{minipage}%
    \begin{minipage}{0.5\textwidth}
        \centering
        \captionof{table}{Data dimensionality} 
        \label{tab:table_DOF}
        \begin{tabular}{lcc} 
            \toprule
            & FOM ($h$) & ROM ($n$) \\ 
            \midrule
            $\bm{u}$ & 1,932 & 1 \\
            $\sigma_\text{vm}$ & 966 & 8 \\
            $\bar{\varepsilon}_\text{pl}$ & 966 & 4 \\
            $\bm{\varepsilon}_\text{pl}$ & 5,796 & 3 \\
            \bottomrule
        \end{tabular}
    \end{minipage}
\end{figure}

\begin{figure}[H]
\centering
\includegraphics[width=1\textwidth]{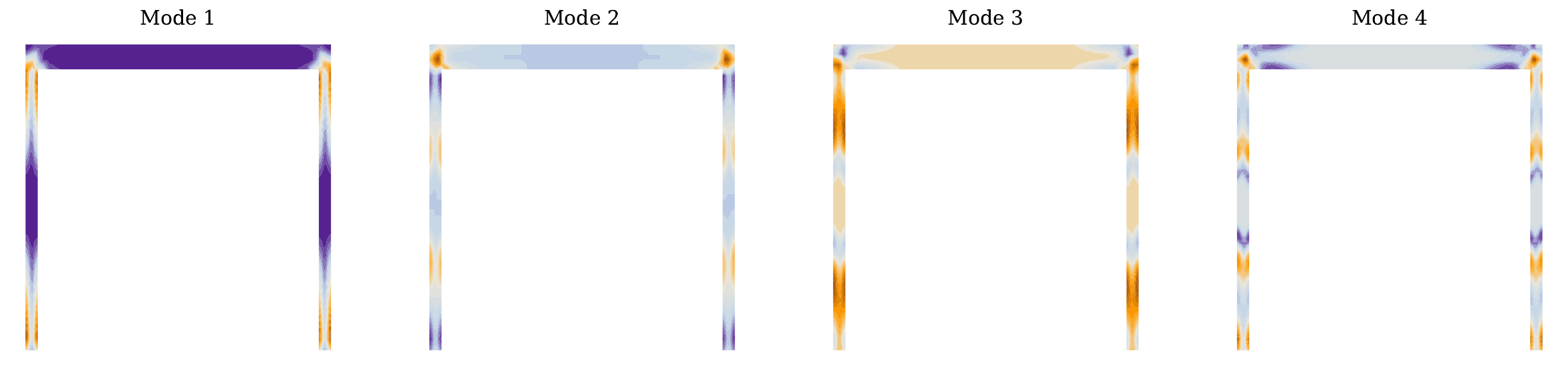}
\captionsetup{justification=centering} 
\caption{The modes of the equivalent plastic strain.}
\label{fig:table_modes}
\end{figure}

By adjusting the accuracy of the POD reduction through variations in $e_{\text{RMSE}\%}$, it is possible to alter \replaced{$n$}{$N_h$}\chglabel{1.13a}, as demonstrated in Figure \ref{fig:POD_table}. This figure illustrates the impact of varying $e_{\text{RMSE}\%}$ values. Table \ref{tab:table_DOF} shows the dimensions, representing $N_h$ for the FOM and the ROM dimensionality, denoted as $n$. This is specifically addressed for a scenario where $e_{\text{RMSE}\%}=1\%$.
\noindent
Figure \ref{fig:table_modes} illustrates the modes $\bm{\Phi}$ related to equivalent plastic strain, showing areas of localized strain.
\replaced{}{Unlike other variables, von Mises stress exhibits significant variation across modes, indicating the need for higher dimensionality to achieve a precise representation.}\chglabel{1.11a}

\subsubsection{Model}
The optimal hyperparameters for the neural networks for this particular dataset, are detailed in Table \ref{tab:table_hyperparameters}. With the selected hyperparameters, the neural network consists of 8058 trainable parameters $\bm{\theta}$\added{, which maps the input of single sample $\mathbf{X} \in \mathbb{R}^{1 \times 10}$ to the output $\mathbf{Y} \in \mathbb{R}^{16 \times 10}$.}\label{comment:1.1} \chglabel{2.1a} 
\begin{table}[h]
    \caption{Optimal hyperparameters.}
    \label{tab:table_hyperparameters}
    \begin{tabular}{p{4cm}p{2cm}}
        \toprule
        Hyperparameter & Value \\
        \midrule
        Batch Size & $32$ \\
        LSTM Hidden State Size & $41$ \\
        Learning Rate & \num{4.5e-2} \\
        LSTM Layers & $1$ \\
        Epochs & $5000$ \\
        Weight Decay & \num{5.4e-4} \\
        \bottomrule
    \end{tabular}
\end{table}
\noindent
\subsubsection{Model Evaluation}
\begin{figure}[H]
    \centering
    \begin{subfigure}{0.48\textwidth}
        \centering
        \includegraphics[width=\textwidth]{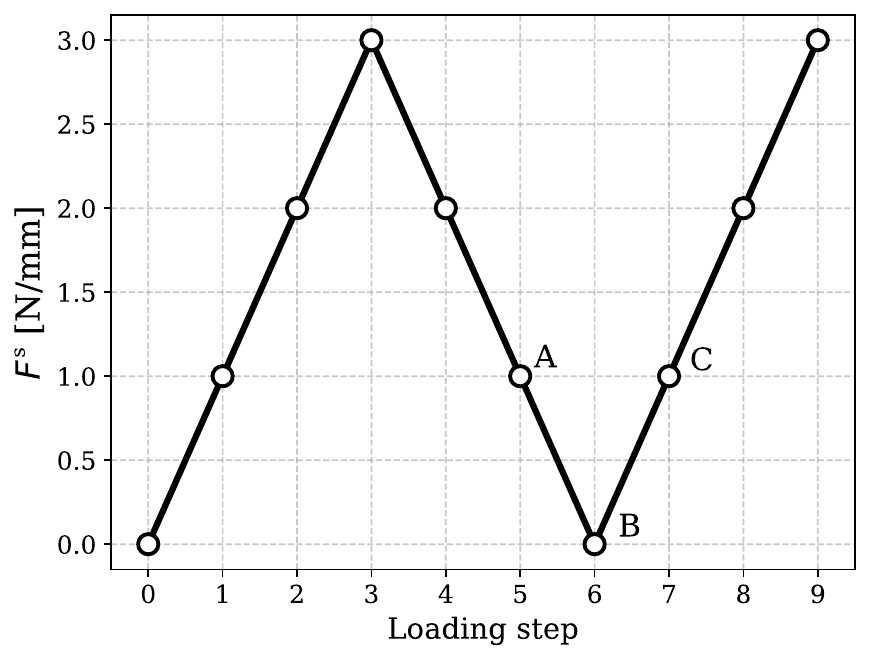}
        \caption{Force history}
        \label{fig:sample3_force_history}
    \end{subfigure}\hfill
    \begin{subfigure}{0.48\textwidth}
        \centering
        \includegraphics[width=\textwidth]{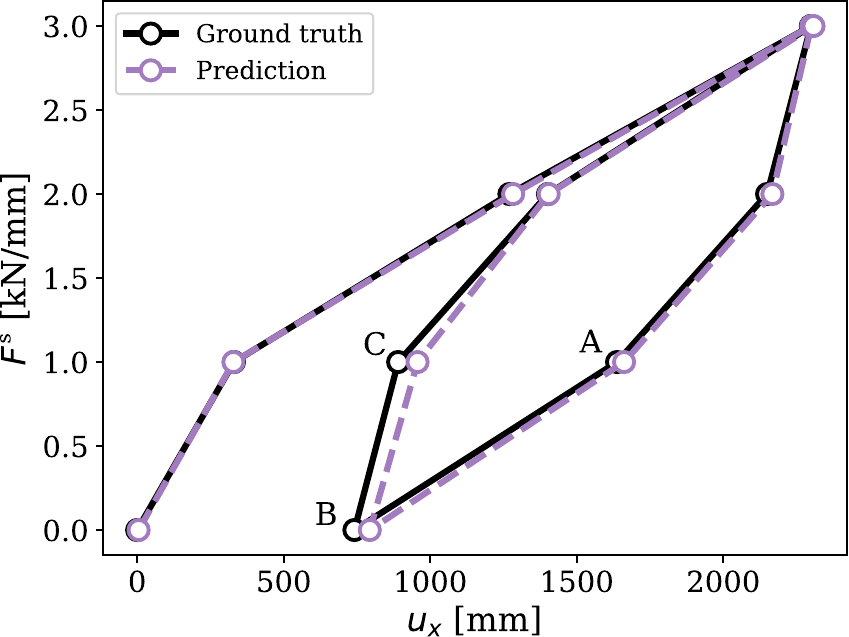}
        \caption{Force-displacement at location L}
        \label{fig:1_sample3_force_state}
    \end{subfigure}
    \caption{Cyclic loading sample.}
    \label{fig:sample3}
\end{figure}

The metrics provided in Table \ref{tab:results_table} demonstrate the framework's accuracy in predicting state variables, with low MAE\% and Weighted $R^2$ across the test set, confirming its ability to model nonlinear elastoplastic behavior and generalize well. 
\begin{table}[h]
    \caption{Multi-task model evaluation.}
    \label{tab:results_table}
    \begin{tabular}{lccc|c}
        \toprule
         & \multicolumn{3}{c}{{$\text{MAE\%}$}} & \multicolumn{1}{c}{$R^2$} \\
        \cmidrule(lr){2-4} \cmidrule(lr){5-5}
        & \multicolumn{1}{|c}{Mean} & Std & Max & Weighted \\
        \midrule
        \multicolumn{1}{c|}{$\bm{u}$}                           & 0.33\%         & 0.84\% & 4.38\% & 1.00     \\
        \multicolumn{1}{c|}{$\sigma_\text{vm}$}           & 0.21\%         & 0.28\% & 2.71\% & 0.96     \\
        \multicolumn{1}{c|}{$\bar{\varepsilon}_\text{pl}$} & 0.30\%         & 0.66\% & 5.04\% & 0.98     \\
        \multicolumn{1}{c|}{$\bm{\varepsilon}_\text{pl}$ }      & 0.10\%         & 0.36\% & 4.97\% & 0.99     \\
        \bottomrule
    \end{tabular}
\end{table}
\newline
The findings are consistent with \citet{Im2021SurrogateDecomposition}, who reported MAE between 0.88\% and 1.09\% in deformation predictions for their test set. The weighted $R^2$ values, nearing 1, indicate the neural network's ability to capture the overall variation.
Given the significance of kinematic hardening and isotropy, particular attention is devoted to a cyclic loading case (sample 3). Figure \ref{fig:sample3_force_history} shows the shear force applied at every loading step for sample 3 and its predicted deformation in the x direction, $u_x$, with a focus on three specific reference steps during unloading.

\begin{figure}[H]
    \centering
    \includegraphics[width=1.\textwidth]{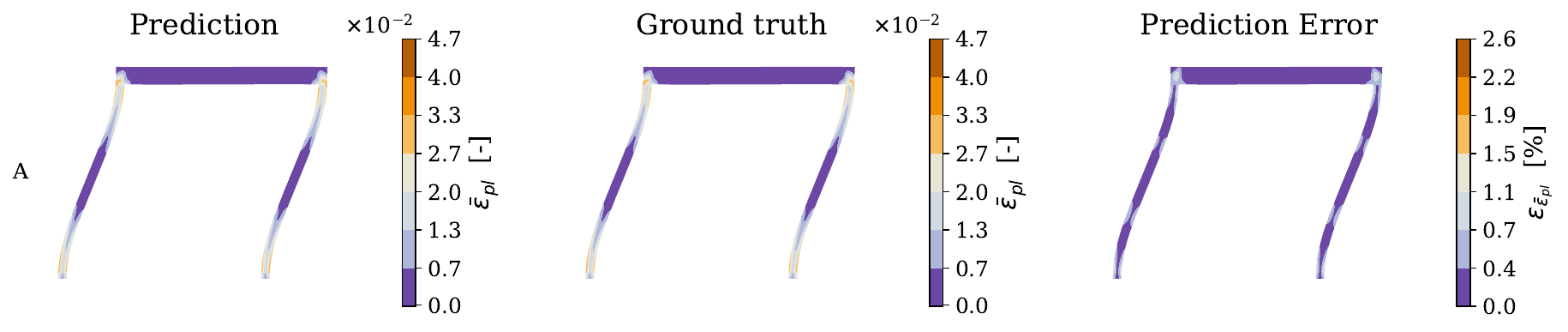}
    \includegraphics[width=1\textwidth]{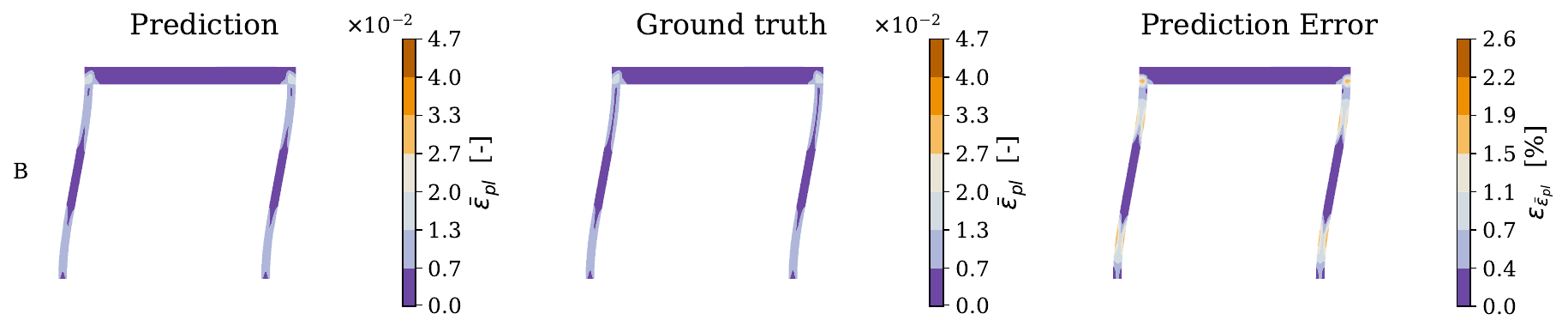}
    \includegraphics[width=1\textwidth]{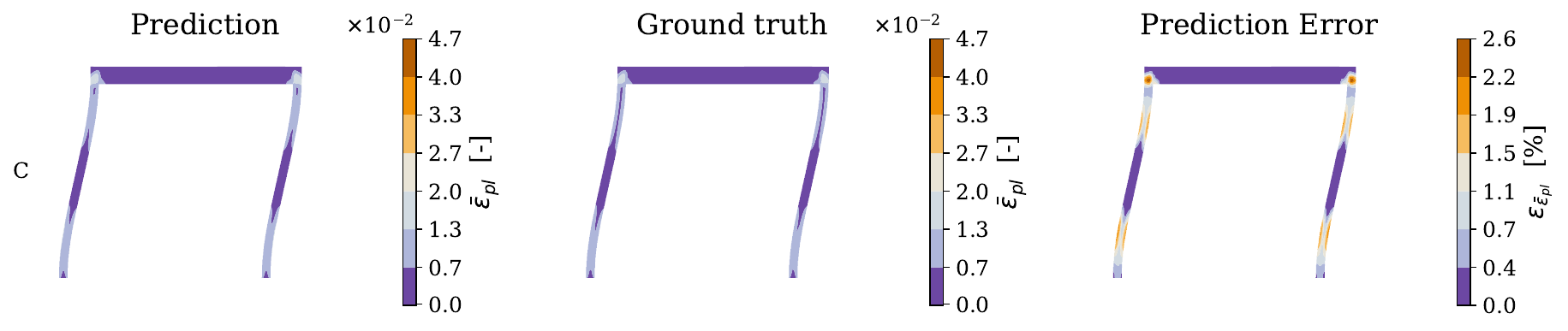}
    \caption{Equivalent plastic strain at each reference point.}
    \label{fig:contour_3}
\end{figure}

\vspace{-0.5cm}
\noindent
The model's effectiveness in capturing material behavior is illustrated by the force-displacement curve at point L, presented in Figure \replaced{\ref{fig:1_sample3_force_state}}{\ref{fig:2_sample3:force_state}}\chglabel{1.12a}, with point L highlighted in Figure \ref{fig:table_geom}. However, deviations are notable at loading steps B and C. Figure \ref{fig:contour_3} further illustrates the equivalent plastic strain, revealing errors mainly around the leg joints. This area shows the greatest magnitude among all modes, significantly affecting predictions due to its critical role.

\vspace{-0.5cm}

\subsubsection{Multi-Task Versus Single-Task}
\begin{figure}[H]
    \centering
    \begin{subfigure}{0.48\textwidth}
        \centering
        \includegraphics[width=\textwidth]{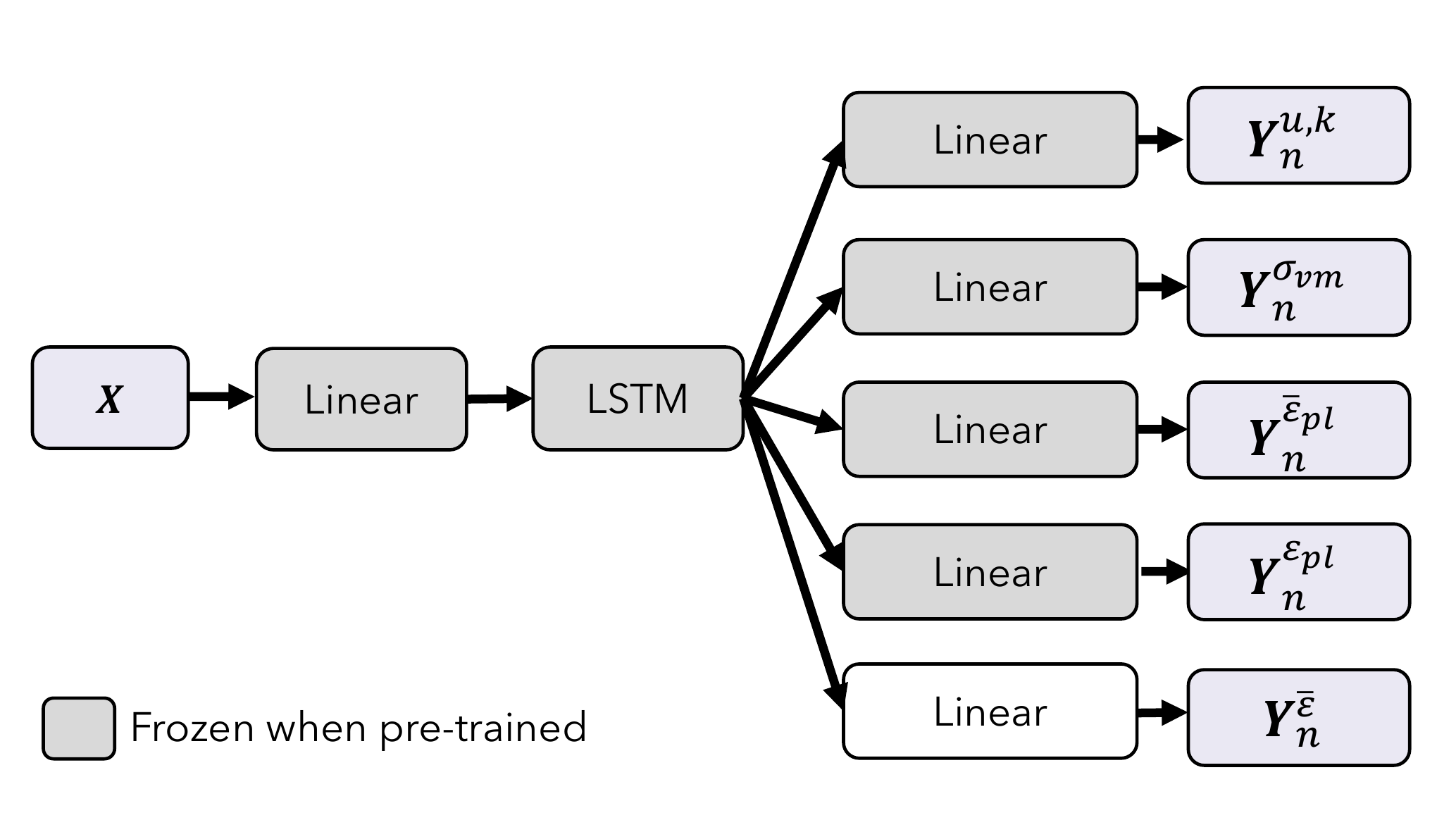}
        \caption{Multi-task networks}
        \label{fig:MTL}
    \end{subfigure}\hfill
    \begin{subfigure}{0.48\textwidth}
        \centering
        \includegraphics[width=\textwidth]{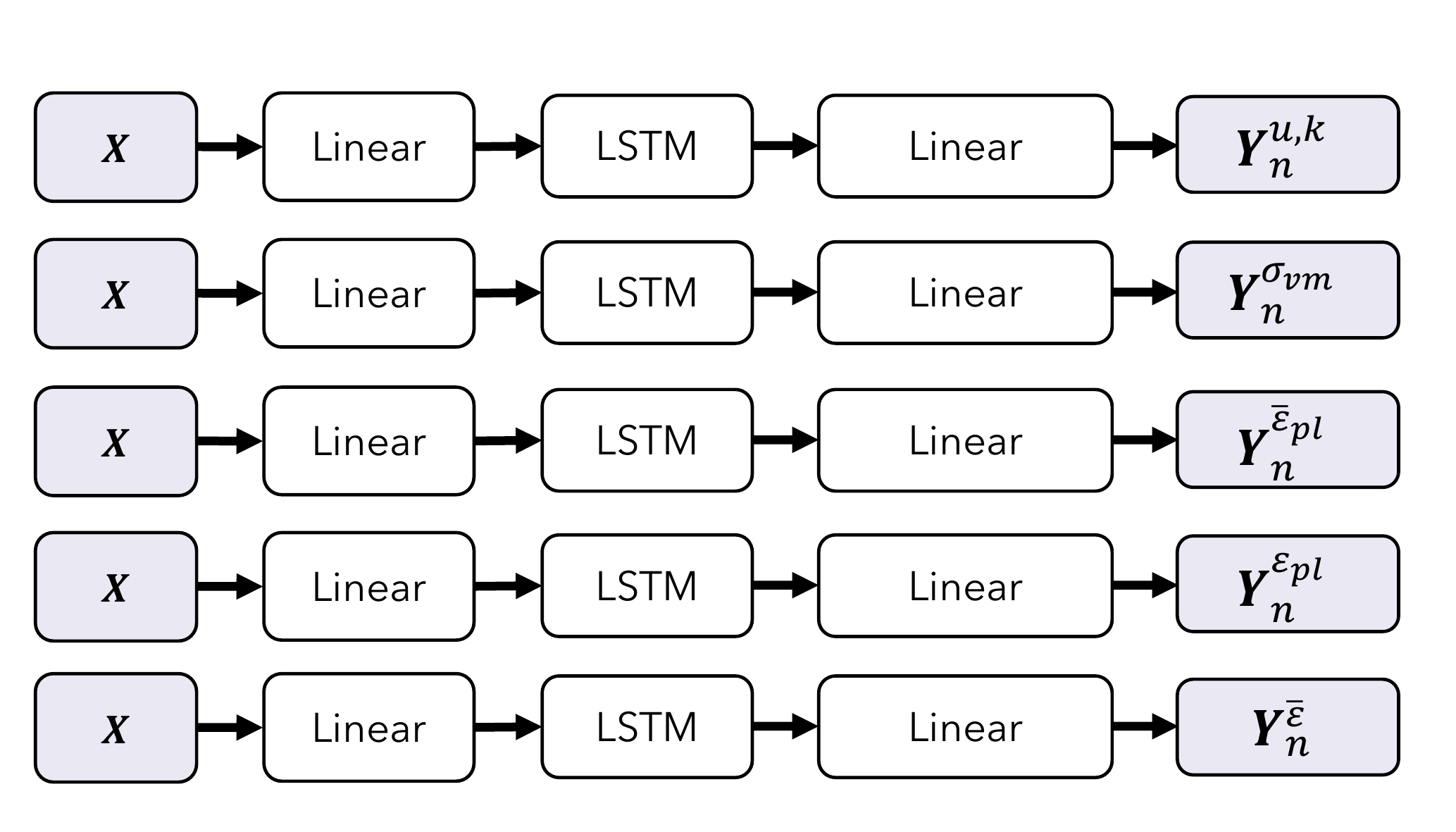}
        \caption{Single-task networks}
        \label{fig:STL}
    \end{subfigure}
    \caption{Network Architectures}
    \label{fig:MTL_VS_STL}
\end{figure}

Prediction errors during cyclic loading at steps B and C hint at potential deviations in the shared layers' representation of the problem. On the other hand, the lack of a shared representation in single-task models might cause larger errors. \added{The models under consideration share the same input $\mathbf{X}$ and hidden state dimensions, but differ in their output dimensions, $\mathbb{R}^{n \times 10}$}\added{, as illustrated in Figure \ref{fig:MTL_VS_STL}.}\chglabel{1.3a}\chglabel{2.1b}\chglabel{2.2c} 

\noindent
\added{The test results of the single-task networks, trained and tested on the same datasets}\chglabel{1.3b} as the multi-task network, \added{are presented in Table \ref{tab:Single1}.}\chglabel{1.10a}
\begin{table}[h]
    \centering
    \caption{Single Task Results.}
    \label{tab:Single1}
    \begin{tabular}{lccccccc|c}  
        \toprule
        &  & \multicolumn{6}{c}{$\text{MAE\%}$} & $R^2$ \\
        \cmidrule(lr){3-8} \cmidrule(lr){9-9}
         & \multicolumn{1}{c|}{\added{$n$}\chglabel{1.1a}} & \multicolumn{2}{c}{Mean} & \multicolumn{2}{c}{Std} & \multicolumn{2}{c|}{Max} & Weighted \\
        \midrule
        $\bm{u}$                      & \multicolumn{1}{c|}{1}  & 0.57\% &  & 1.44\% &  & 8.46\% &  & 1.00 \\
        $\sigma_\text{vm}$            & \multicolumn{1}{c|}{8}  & 0.24\% &  & 0.35\% &  & 5.63\% &  & 0.96 \\
        $\bar{\varepsilon}_\text{pl}$ & \multicolumn{1}{c|}{4}  & 0.38\% &  & 0.96\% &  & 8.81\% &  & 0.97 \\
        $\bm{\varepsilon}_\text{pl}$  & \multicolumn{1}{c|}{3}  & 0.16\% &  & 0.66\% &  & 11.75\% &  & 0.99 \\
        \bottomrule
    \end{tabular}
\end{table}
\added{Comparing these to the multi-task results, shown in Table \ref{tab:Single_vs_multi_1}}\chglabel{1.10b}, reveals that the MTL network outperforms the single-task models, depending on the state variable. Here, negative error values and positive weighted $R^2$ scores indicate enhancements over the single-task network.

\begin{table}[h]
    \centering
    \caption{Multi Task Improvements.}
    \label{tab:Single_vs_multi_1}
    \begin{tabular}{lcccccc|c}  
        \toprule
        \multicolumn{1}{c}{} & \multicolumn{6}{c}{$\text{MAE\%}$} & $R^2$ \\
        \cmidrule(lr){2-7} \cmidrule(lr){8-8}
        \multicolumn{1}{c|}{} & \multicolumn{2}{c}{Mean} & \multicolumn{2}{c}{Std} & \multicolumn{2}{c|}{Max} & Weighted \\
        \midrule
        \multicolumn{1}{c|}{$\bm{u}$}                      & -0.24\% &  & -0.60\% &  & -4.08\% &  & 0.00 \\
        \multicolumn{1}{c|}{$\sigma_\text{vm}$}            & -0.03\% &  & -0.07\% &  & -2.92\% &  & 0.00 \\
        \multicolumn{1}{c|}{$\bar{\varepsilon}_\text{pl}$} & -0.08\% &  & -0.30\% &  & -3.77\% &  & +0.01 \\
        \multicolumn{1}{c|}{$\bm{\varepsilon}_\text{pl}$}  & -0.06\% &  & -0.30\% &  & -6.78\% &  & 0.00 \\
        \bottomrule
    \end{tabular}
\end{table}

\begin{figure}[H]
    \centering
    \begin{subfigure}{0.48\textwidth}
        \centering
        \includegraphics[width=\textwidth]{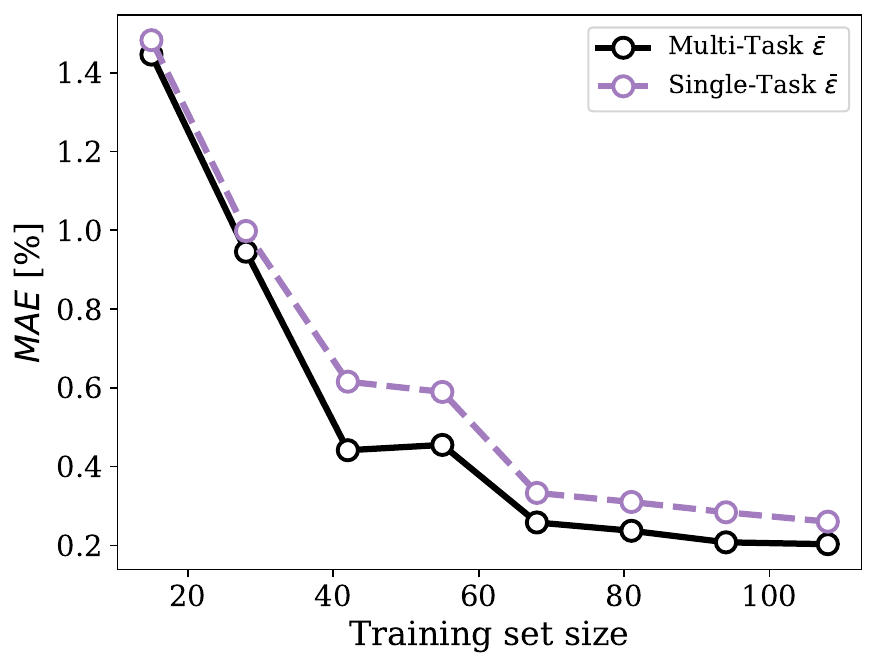}
        \caption{Mean absolute error}
        \label{fig:Overfit_MAE}
    \end{subfigure}\hfill
    \begin{subfigure}{0.48\textwidth}
        \centering
        \includegraphics[width=\textwidth]{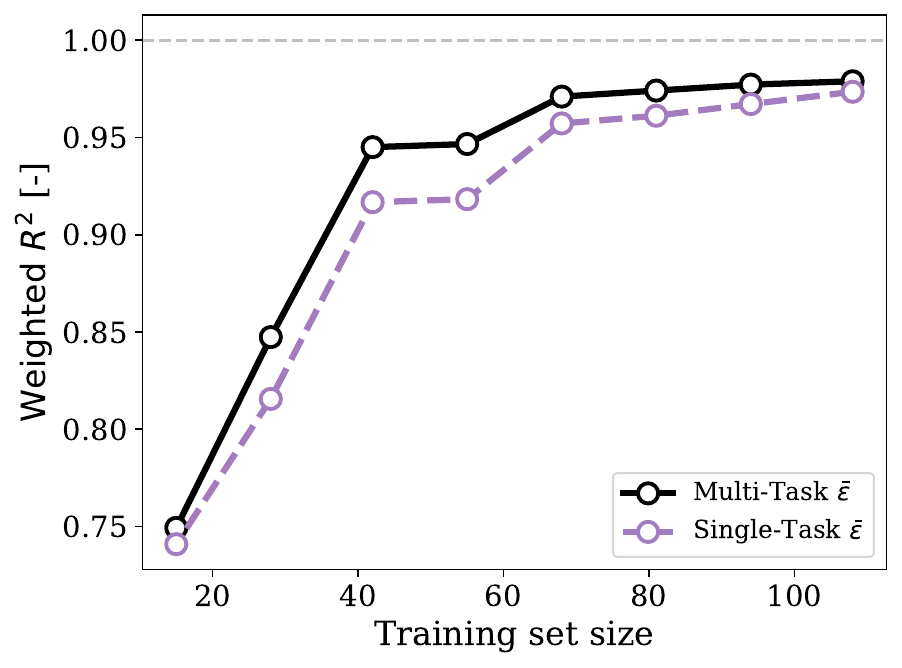}
        \caption{Weighted coefficient of determination}
        \label{fig:Overfit_R2}
    \end{subfigure}
    \caption{Validation set: model comparison for different training set sizes.}
    \label{fig:overfitting}
\end{figure}

\noindent
It is crucial to emphasize that while the mean value differences might seem modest, the multi-task model consistently surpasses the single-task model across all field variables. Notably, the most prominent difference is observed in the maximum values of the displacement, where outliers are significantly reduced. This improvement is especially evident for the displacement field, characterized by $n=1$.\added{Training on a smaller dataset and evaluating on the validation set yields the results shown in Figure \ref{fig:overfitting}. This figure clearly indicates that the single-task model is more susceptible to overfitting, as reflected by its poorer performance on the validation set. In contrast, the multi-task model demonstrates stronger generalization, likely due to its shared layers that help mitigate overfitting. This evaluation was conducted after training stalled for a fixed number of epochs.}\chglabel{1.6a}

\begin{figure}[H]
  \centering
  \begin{subfigure}{0.49\textwidth}
    \centering
    \includegraphics[width=0.92\textwidth]{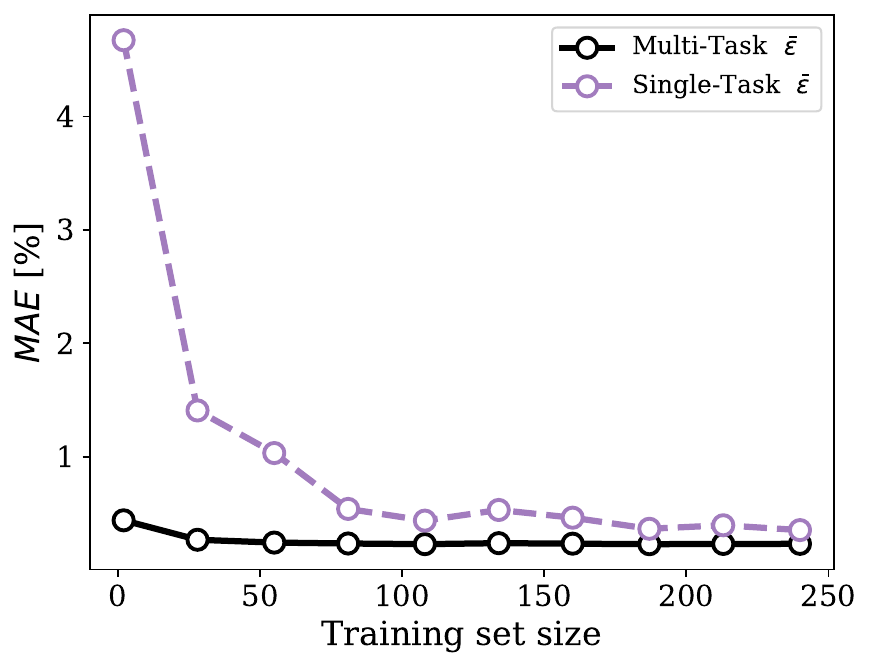}
    \caption{Mean absolute error}
    \label{fig:Sensitivity_eqtot_table}
  \end{subfigure}
  \hfill
  \begin{subfigure}{0.49\textwidth}
    \centering
    \includegraphics[width=0.92\textwidth]{Table_setsize_MAE.pdf}
    \caption{Weighted coefficient of determination}
    \label{fig:Sensitivity_eqtot_beam}
  \end{subfigure}
  \caption{Test set: Model comparison on $\bar{\varepsilon}$ across different training set sizes.}
  \label{fig:eqtot_sensitivity_1}
\end{figure}

\noindent
Moreover, a pre-trained multi-task model can effectively train additional variables with fewer samples due to its comprehensive understanding. For instance, in the case of equivalent strain $\bar{\varepsilon}$ (encompassing both elastic and plastic strain), less than 25 samples are needed to reach an $R^2$ level similar to that achieved with 100 samples in a single-task setup, as shown in Figure \ref{fig:eqtot_sensitivity_1}.
\noindent
 \added{In this approach, we freeze the weights for the pre-trained multi-task model as depicted in Figure \ref{fig:MTL}}\chglabel{1.3c}\chglabel{2.2ba}, \added{with output $\mathbf{Y} \in \mathbb{R}^{(16 + 5)\times 10}$. The first 16 dimensions correspond to previously trained field variables, while the last 5 dimensions capture the reduced-order data for the equivalent strain. In contrast, the single-task neural network has an output $\mathbf{Y} \in \mathbb{R}^{5\times 10}$, where only the equivalent strain is predicted without the benefit of shared knowledge across tasks, resulting in a slower convergence and requiring more training data.}\chglabel{1.1b}\chglabel{2.1c} 

\begin{figure}[H]
    \centering
    \begin{subfigure}{0.48\textwidth}
        \centering
        \includegraphics[width=\textwidth]{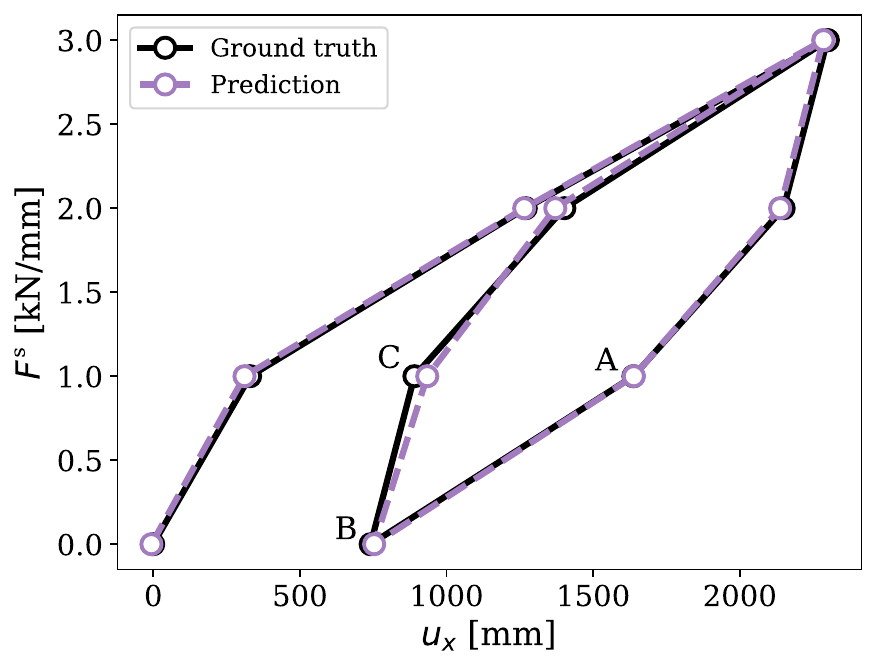}
        \caption{Improved force-displacement at point L}
        \label{fig:sample3_force_state_improved}
    \end{subfigure}\hfill
    \begin{subfigure}{0.48\textwidth}
        \centering
        \includegraphics[width=\textwidth]{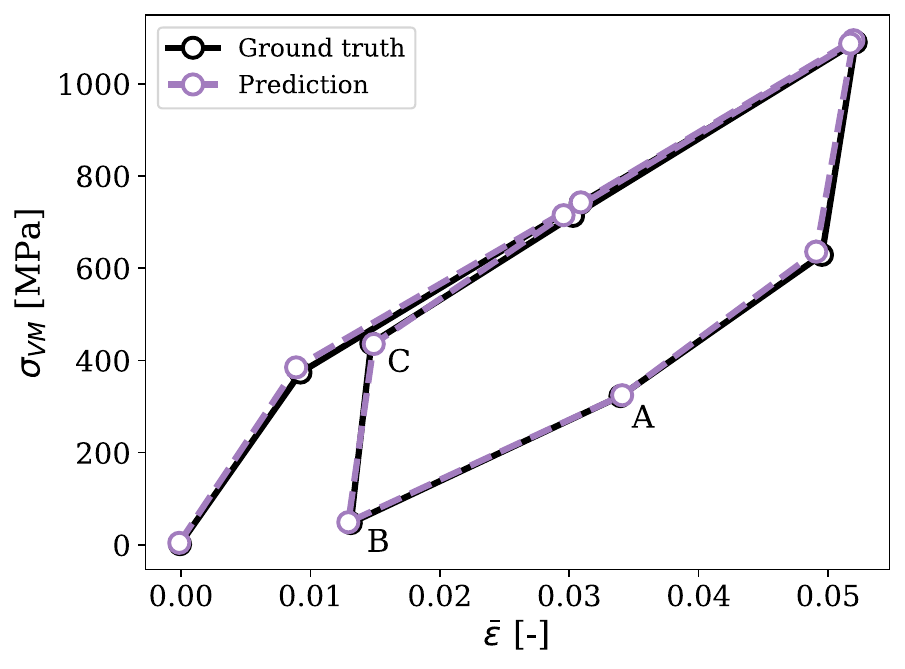}
        \caption{Stress-strain curve at point K}
        \label{fig:2_sample3:force_state}
    \end{subfigure}
    \caption{Cyclic loading sample.}
    \label{fig:sample3_improved}
\end{figure}

Subsequently, incorporating $\bar{\varepsilon}$ \added{and not freezing the variables, but instead allowing them to be trained alongside the additional variable}\chglabel{1.3d}\chglabel{2.2bb}\added{, expands the output dimension to $\mathbf{Y} \in \mathbb{R}^{21 \times 10}$}\chglabel{1.1c}\chglabel{2.1d} . \added{This further boosts performance, as shown by the improvements in Table \ref{tab:extra_variable_table}}
\chglabel{1.14a}
\begin{table}[h]
\centering
    \caption{Enhancements of extra variable $\bar{\varepsilon}$.}
    \label{tab:extra_variable_table}
    \begin{tabular}{lccc|c}  
        \toprule
        \multicolumn{1}{c}{} & \multicolumn{3}{c}{{MAE\%}} & \multicolumn{1}{c}{$R^2$} \\
        \cmidrule(lr){2-4} \cmidrule(lr){5-5}
        \multicolumn{1}{c|}{} & Mean & Std & Max & Weighted \\
        \midrule
        \multicolumn{1}{c|}{$\bm{u}$}                           & -0.06\%          & -0.28\% & -1.00\% & 0.00        \\
        \multicolumn{1}{c|}{$\sigma_\text{vm}$}            & -0.01\%           & 0.00\% & +0.49\% & +0.01     \\
        \multicolumn{1}{c|}{$\bar{\varepsilon}_\text{pl}$} & -0.05\%          & -0.14\% & +0.26\%  & +0.01     \\
        \multicolumn{1}{c|}{$\bm{\varepsilon}_\text{pl}$}       & -0.02\%          & -0.08\% & -0.18\% & 0.00    \\
        \bottomrule
    \end{tabular}
\end{table}
\noindent
\added{These results indicate an overall improvement, except for the maximum values of both $\sigma_\text{vm}$ and $\bar{\varepsilon}_\text{pl}$.}\chglabel{1.14b} Notably, \replaced{load step}{time step}\chglabel{2.4gi} B and C align more closely with the ground truth, as shown in Figure \ref{fig:sample3_force_state_improved}. In the case of the stress-strain curve, shown in Figure \ref{fig:2_sample3:force_state}, both stress and strain closely match the ground truth, demonstrating the model’s capability to accurately replicate the Bauschinger effect in the shared layers.

\vspace{-0.25cm}
\subsection{Use case 2: Cantilever-Beam}\label{subsec2}
\begin{figure}[H]
        \centering
        \includegraphics[width=1\linewidth]{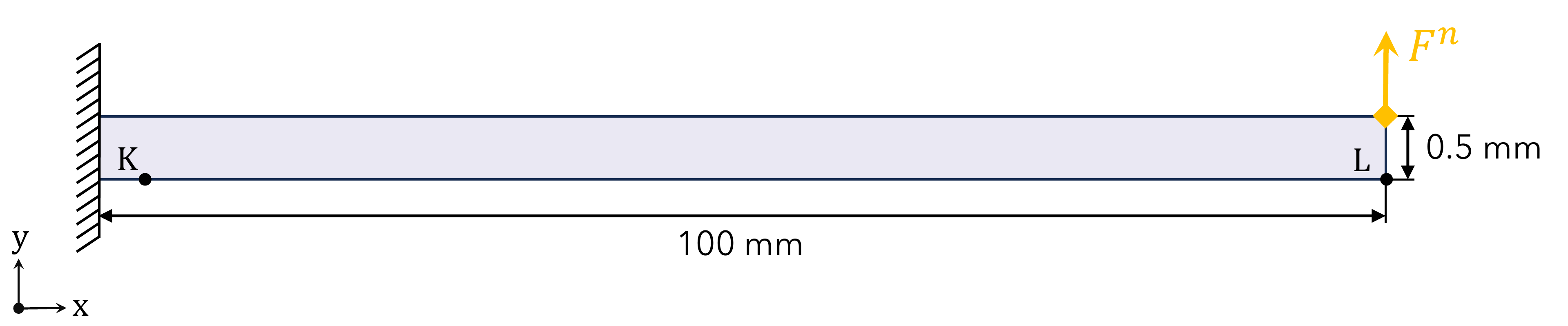}
        \caption{Beam model.}
        \label{fig:beam_geom}
\end{figure}

This section evaluates the performance of the proposed LSTM MTL framework through a 2D cantilever problem, as illustrated in Figure \ref{fig:beam_geom}. The cantilever problem is selected for analysis because it serves as a simplified model of a catheter, which can be used, for example, for surgical assistance.

\noindent
The FEM model consists of 980 linear 2D quadrilateral elements and 1128 nodes. Data for field variables were created by applying a diverse set of normal point forces across $N_t=10$ \replaced{load step}{time step}\chglabel{2.4gj} and adjusting the force's position atop the beam. The parameter set $\mu$, covering all boundary conditions, includes $N_\mu=2$ with a domain of $\mathcal{P} = [-6,6] \text{ N} \times [5,10] \text{ mm}$. Here, the first interval specifies the magnitude of the normal force $F^n$, and the second indicates its x-direction position $P_x$.

For testing, positions were fixed at $P_x=7.5$ mm, except in the constant force sample, where the position shifts from the beam's center to its end, isolating the additional input parameter's effect in this specific sample.

\subsubsection{Data Reduction}
\begin{figure}[H]
    \centering
    \begin{minipage}{0.5\textwidth}
        \centering
        \includegraphics[width=\linewidth]{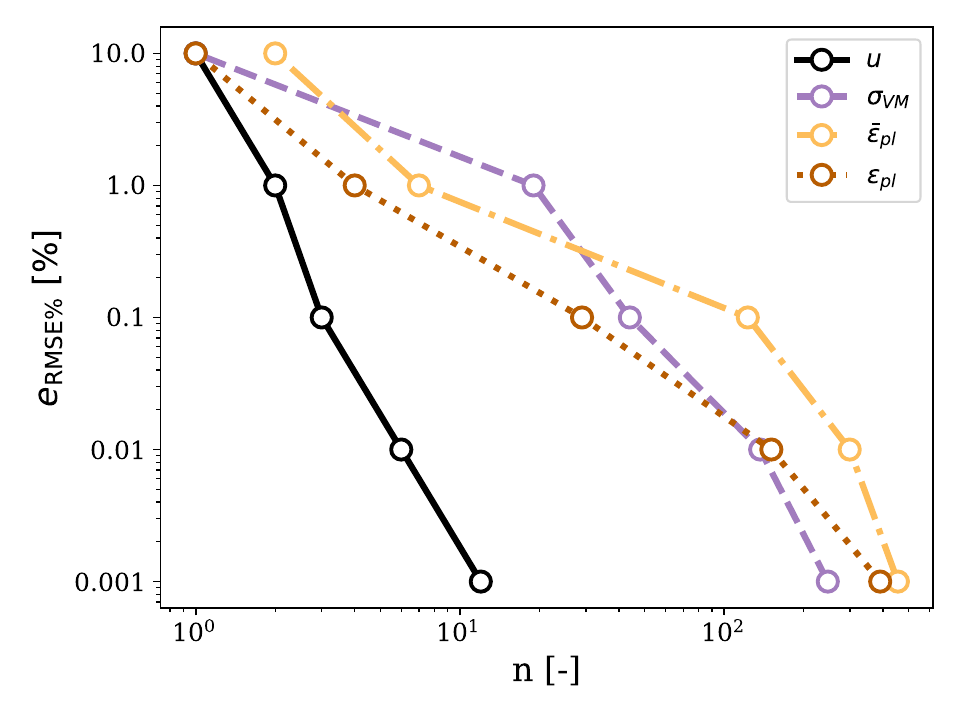}
        \caption{\added{POD accuracy vs efficiency.}}
        \label{fig:POD_beam}
    \end{minipage}%
    \begin{minipage}{0.5\textwidth}
        \centering
        \captionof{table}{Data dimensionality} 
        \label{tab:beam_DOF}
            \begin{tabular}{lcc} 
            \toprule
            & FOM ($N_h$) & ROM ($n$) \\
            \midrule
            $\bm{u}$                  & 2,256          & 2 \\
            $\sigma_\text{vm}$   & 1,128          & 19 \\
            $\bar{\varepsilon}_\text{pl}$ & 1,128    & 7 \\
            $\bm{\varepsilon}_\text{pl}$  & 6,768          & 4 \\
            \bottomrule
            \end{tabular}
    \end{minipage}
\end{figure}

After the data collection, a detailed description of the dimensionality reduction applied to the data can be found in Table \ref{tab:beam_DOF} for $e_{\text{RMSE}\%}=1\%$ and for a varying $e_{\text{RMSE}\%}$ in Figure \ref{fig:POD_beam}.

\begin{figure}[H]
    \centering
    \includegraphics[width=\linewidth]{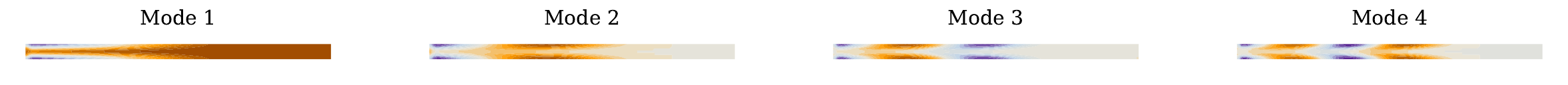}
    \captionsetup{justification=centering} 
    \caption{The first 4 modes of the equivalent plastic strain.}
    \label{fig:beam_modes}
\end{figure}

\noindent
In contrast to use case 1, the variables need to be represented with a greater number of modes and eigenvalues. This is due to the minor differences in the modes, mainly caused by the changing $P_x$ leading to more variability in local areas, and resulting in a wider variety of loaded regions. For the equivalent plastic strain, the first modes are depicted in Figure \ref{fig:beam_modes}.

\subsubsection{Model}
The optimal hyperparameters obtained via the search are displayed in Table \ref{tab:beam_hyperparameters}, resulting in a total of 30132 trainable parameters $\bm{\Phi}$ \added{, which maps the input for a single sample $\mathbf{X} \in \mathbb{R}^{2 \times 10}$ to the output $\mathbf{Y} \in \mathbb{R}^{35 \times 10}$}\chglabel{1.1d}\chglabel{2.1e} . 
\begin{table}[h]
    \centering
    \caption{Optimal hyperparameters.}
    \label{tab:beam_hyperparameters}
    \begin{tabular}{p{4cm}p{2cm}}
        \toprule
        Hyperparameter & Value \\
        \midrule
        Batch Size & $71$ \\
        LSTM Hidden State Size & $80$ \\
        Learning Rate & \num{3.3e-2} \\
        Number of LSTM Layers & $1$ \\
        Epochs & $1000$ \\
        Weight Decay & $1.5 \times 10^{-6}$ \\
        \bottomrule
    \end{tabular}
\end{table}
\subsubsection{Model Evaluation}
Training five unique neural networks with these hyperparameters, followed by averaging their outputs, results in the test outcomes displayed in Table \ref{tab:beam_results}. 
\begin{table}[h]
    \centering
    \caption{Model evaluation.}
    \label{tab:beam_results}
    \begin{tabular}{lccc|c}  
        \toprule
        \multicolumn{1}{c}{} & \multicolumn{3}{c}{{MAE\%}} & \multicolumn{1}{c}{$R^2$} \\
        \cmidrule(lr){2-4} \cmidrule(lr){5-5}
        \multicolumn{1}{c|}{} & Mean & Std & Max & Weighted \\
        \midrule
        \multicolumn{1}{c|}{$\bm{u}$}                           & 0.08\%           & 0.21\% & 2.57\% & 1.00     \\
        \multicolumn{1}{c|}{$\sigma_\text{vm}$}            & 0.27\%           & 0.33\% & 5.73\% & 0.93     \\
        \multicolumn{1}{c|}{$\bar{\varepsilon}_\text{pl}$} & 0.12\%           & 0.29\% & 4.29\% & 0.97     \\
        \multicolumn{1}{c|}{$\bm{\varepsilon}_\text{pl}$}       & 0.05\%           & 0.16\% & 3.29\% & 0.99     \\
        \bottomrule
    \end{tabular}
\end{table}
\newline
\noindent
The results are consistent with the magnitude observed in the prior use case. It is noteworthy that the MAE for the Von Mises stress is relatively higher compared to other variables. 
This increased MAE may be attributed to the necessity for a greater number of output dimensions for this state variable.

\begin{figure}[H]
  \centering
  \begin{subfigure}{0.49\textwidth}
    \centering
    \includegraphics[width=0.9\textwidth]{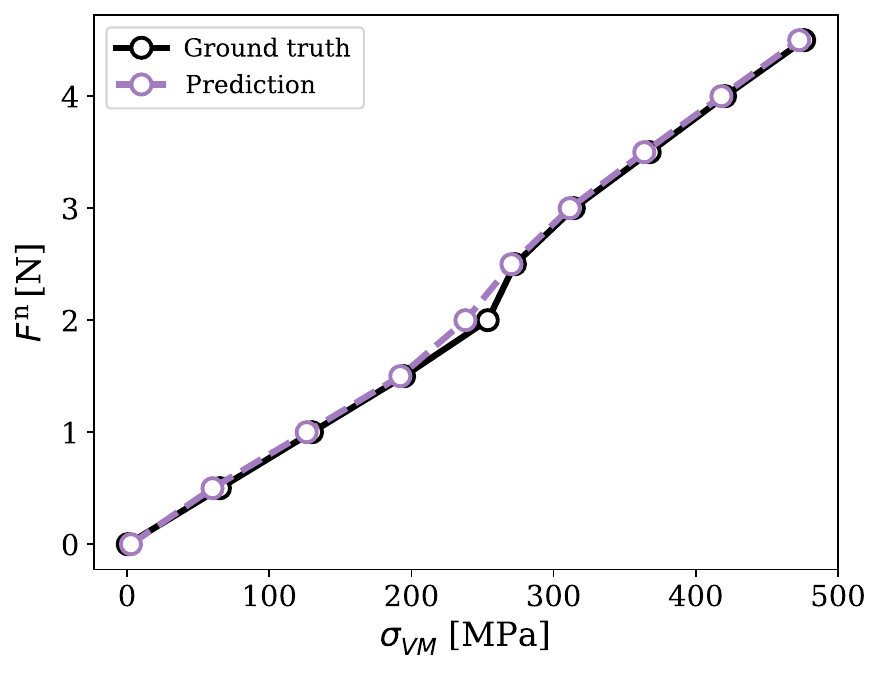}
    \caption{Sample 1: Force-stress}
    \label{fig:force_stress_1}
  \end{subfigure}
  \hfill
  \begin{subfigure}{0.49\textwidth}
    \centering
\includegraphics[width=0.9\textwidth]{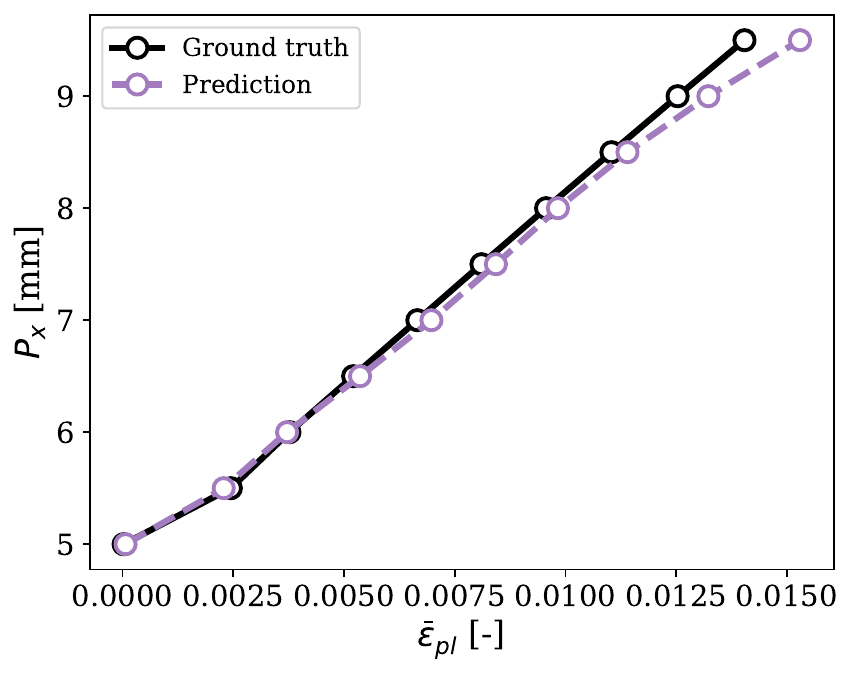}
    \caption{Sample 4: Position-plastic strain}
    \label{fig:position_strain_4}
  \end{subfigure}
  \caption{Influence of varying boundary condition at point K.}
  \label{fig:beam_samples}
\end{figure}

\begin{figure}[H]
  \centering
  \begin{subfigure}{0.49\textwidth}
    \centering
    \includegraphics[width=0.9\textwidth, trim={2cm 1.5cm 0cm 3cm}, clip]{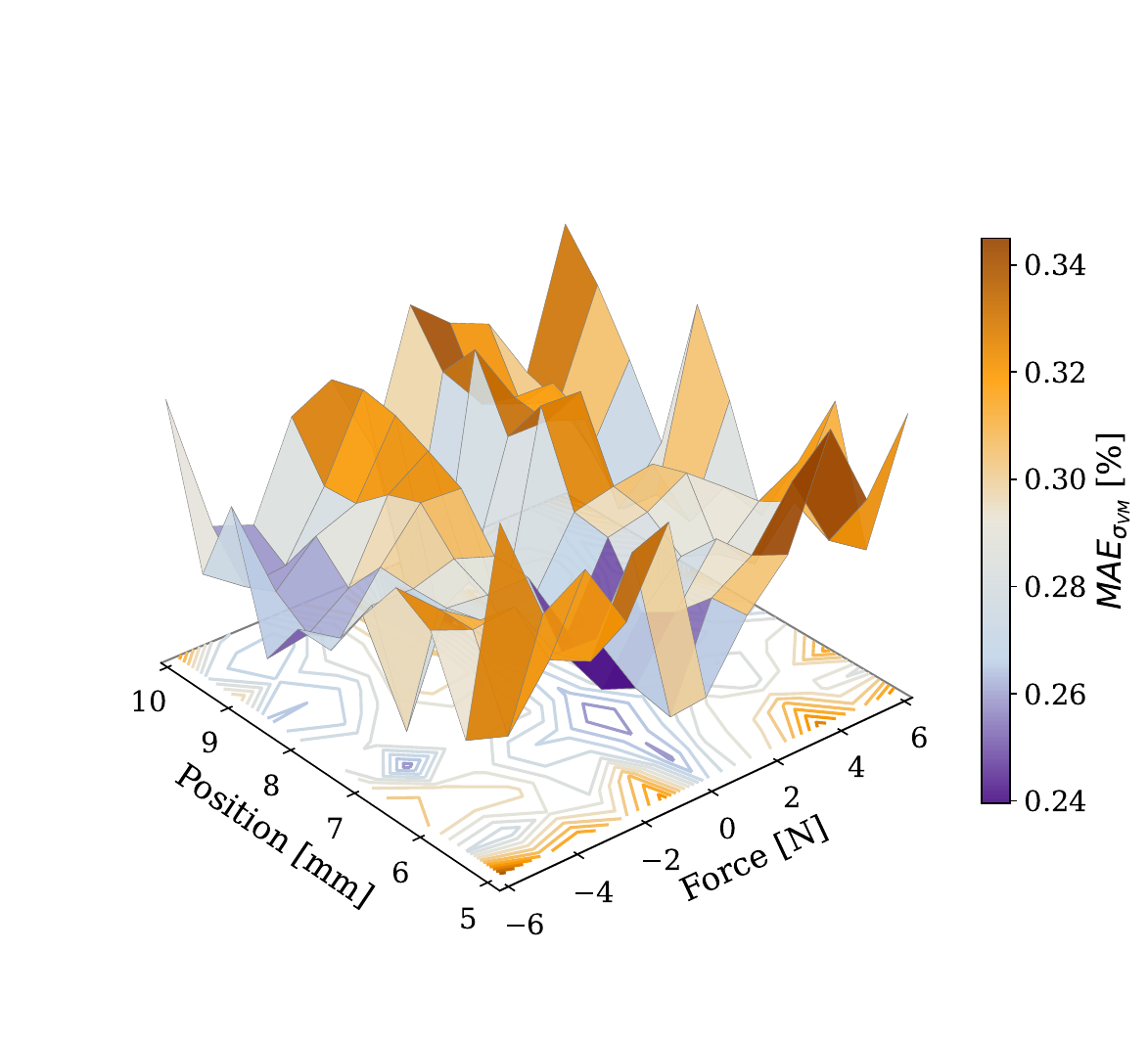}
    \caption{Von Mises stress}
    \label{fig:extrapolate_VM}
  \end{subfigure}
  \hfill
  \begin{subfigure}{0.49\textwidth}
    \centering
    \includegraphics[width=0.9\textwidth, trim={2cm 1.5cm 0cm 3cm}, clip]{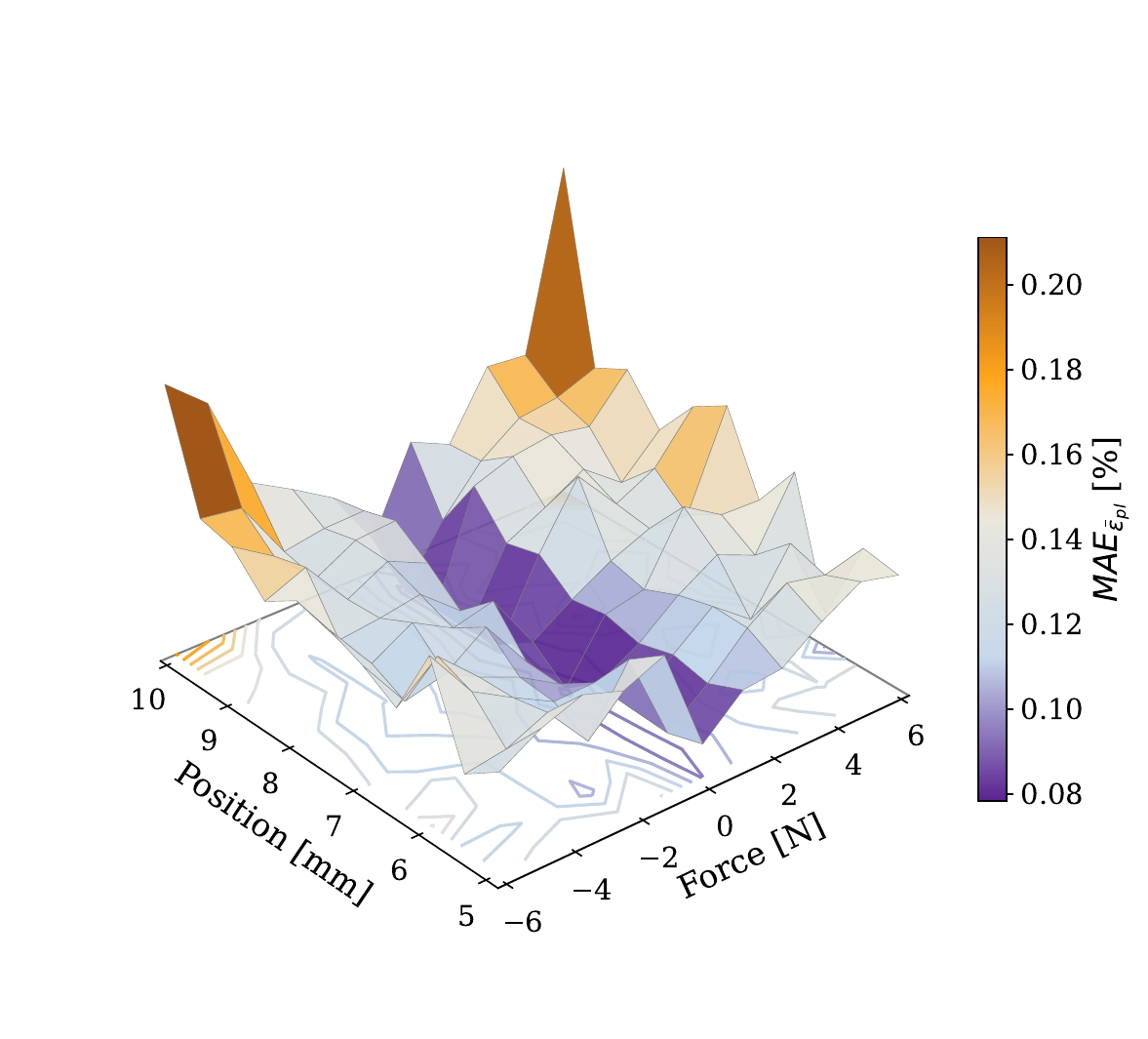}
    \caption{Equivalent plastic strain}
    \label{fig:extrapolate_E}
  \end{subfigure}
  \caption{Training Set: MAE input analysis.}
  \label{fig:extrapolate}
\end{figure}

In the first test sample, where the load increases while the position remains constant, the force-stress curve closely matches the ground truth, as illustrated in Figure \ref{fig:force_stress_1}. However, there is a minor discrepancy during the transition from elastic to plastic behaviour. For the sample where the beam is subjected to a constant force but varying position, there is a noticeable difference in the predicted equivalent plastic strain as the position approaches the beam's end as seen in Figure \ref{fig:position_strain_4}.
\noindent
Limitations are evident in the training set samples, as Figure \ref{fig:extrapolate} demonstrates the model's predictive accuracy, contrasting input variables against average errors at each \replaced{load step}{timestep}\chglabel{2.4gk}. For the initial sample, noticeable errors around the yield point are highlighted in Figure \ref{fig:extrapolate_VM}, where near-zero forces result in a minimal error, increasing with forces inducing plastic flow (around $F^n=1$ N). This error spike reflects the material's differing behavior in the elastic versus plastic regions. Equivalent plastic strain issues, potentially significant, are lessened as transitions from zero to minor strains prevent large MAE\% peaks, as Figure \ref{fig:extrapolate_E} displays.

\begin{figure}[H]
    \centering
    \includegraphics[width=0.50\linewidth]{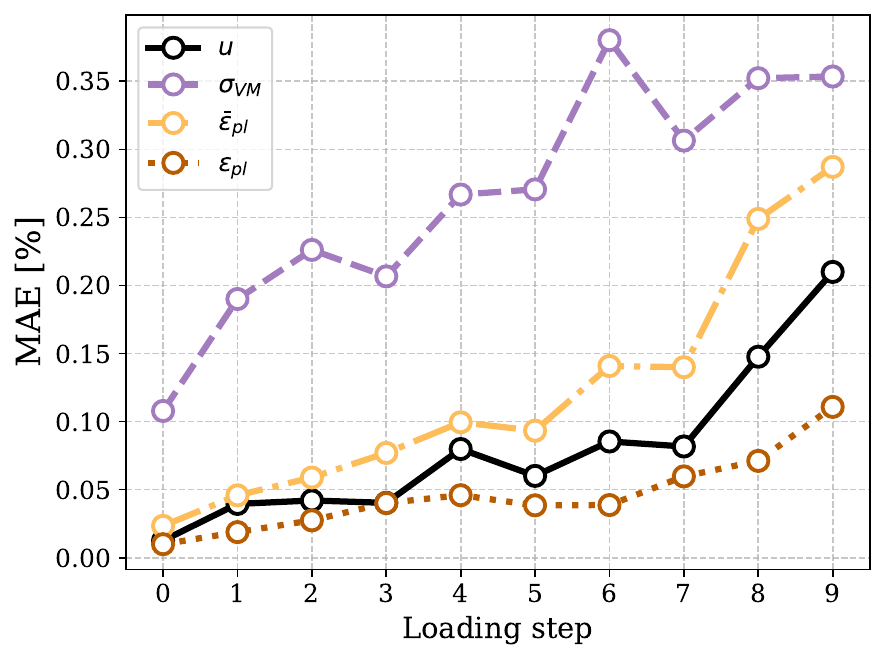}
    \captionsetup{justification=centering} 
    \caption{Compound error.}
    \label{fig:Error_time}
\end{figure}

Moreover, the fourth sample shows error amplification towards the beam's end, illustrating the neural network's difficulty with predictions near input range boundaries due to data scarcity in those areas, a challenge also seen near limits in Figure \ref{fig:extrapolate_VM}.
\noindent
Therafter, the networks error increases over time due to its LSTM architecture, as shown in Figure \ref{fig:Error_time} for the entire test set. This rising error for each variable is likely caused by the LSTM's sequential processing, where initial inaccuracies can compound, leading to greater errors in later predictions. This compound error effect, where small inaccuracies in earlier predictions amplify over time, becomes particularly significant in long-duration simulations or scenarios with rapidly changing variables.

\begin{figure}[H]
    \centering
    \begin{subfigure}{0.49\textwidth}
        \includegraphics[width=1\linewidth]{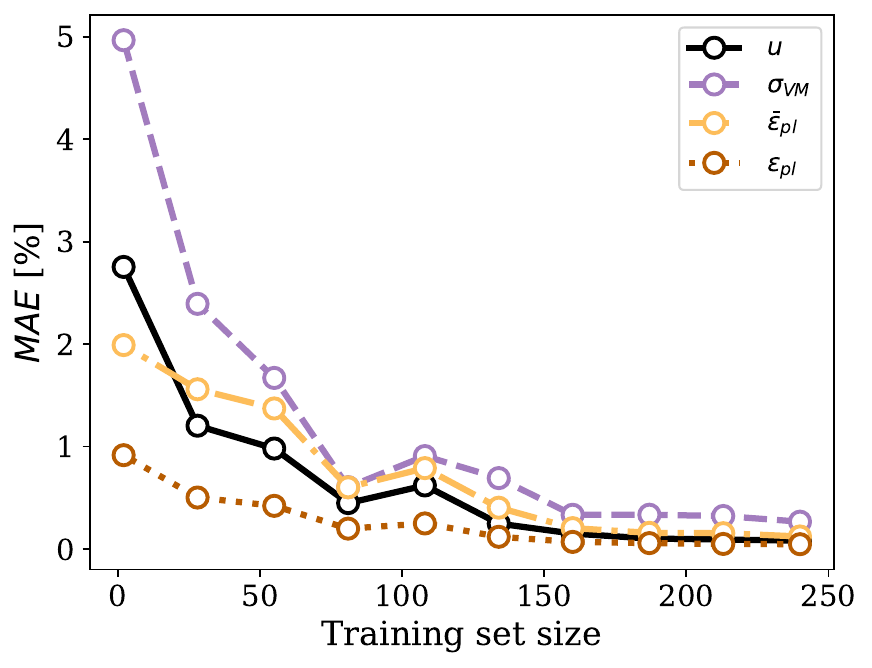}
        \caption{Relative Mean absolute error}
        \label{fig:subfig1}
    \end{subfigure}%
    \hfill
    \begin{subfigure}{0.49\textwidth}
        \includegraphics[width=1\linewidth]{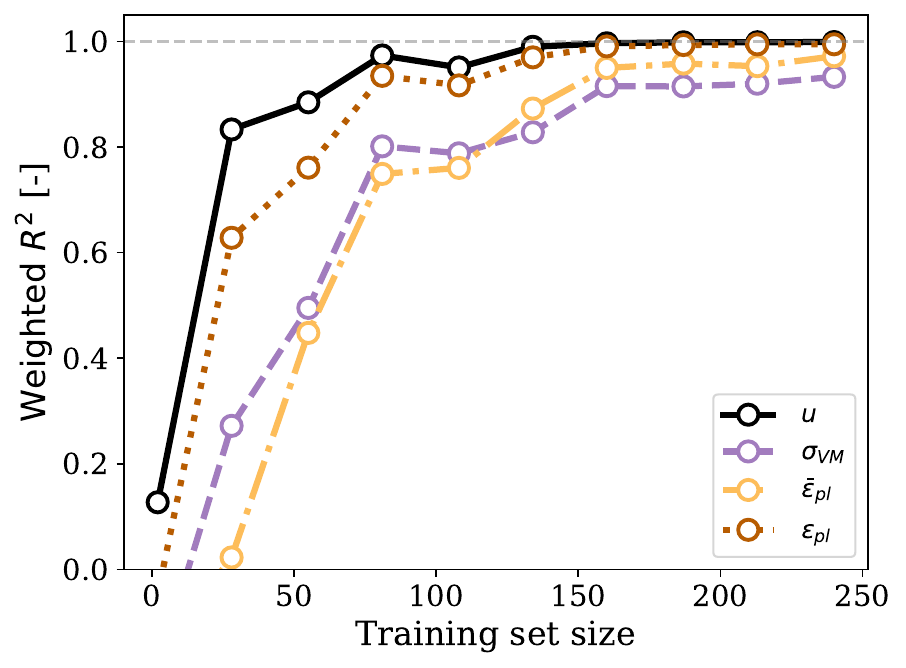}
        \caption{Weighted coefficient of determination}
        \label{fig:subfig2}
    \end{subfigure}%
    \caption{Test set results with varying training set size.}
    \label{fig:Beam_data_sens}
\end{figure}

\noindent
Reducing the training set size, as shown in Figure \ref{fig:Beam_data_sens}, reveals that fewer than 170 samples are needed for MAE\% and weighted $R^2$ convergence. 

\subsubsection{Multi-Task Versus Single-Task}
Table \ref{tab:Single2} presents the test results for the single-task networks, \added{based on the same datasets}\chglabel{1.3e}, \added{ while a comparison with the multi-task network, is shown in Table \ref{tab:Single_vs_multi_2}}\chglabel{1.10c}

\begin{table}[h]
    \caption{Single Task Results.}
    \label{tab:Single2}
    \centering
    \begin{tabular}{lccccccc|c}
        \toprule
        &  & \multicolumn{6}{c}{$\text{MAE\%}$} & $R^2$ \\
        \cmidrule(lr){3-8} \cmidrule(lr){9-9}
         & \multicolumn{1}{c|}{\added{$n$}\label{1.1e}} & \multicolumn{2}{c}{Mean} & \multicolumn{2}{c}{Std} & \multicolumn{2}{c|}{Max} & Weighted \\
        \midrule
        $\bm{u}$                      & \multicolumn{1}{c|}{2} & 0.12\% &  & 0.25\% &  & 2.66\% &  & 1.00 \\
        $\sigma_\text{vm}$            & \multicolumn{1}{c|}{19}  & 0.23\% &  & 0.35\% &  & 5.86\% &  & 0.90 \\
        $\bar{\varepsilon}_\text{pl}$ & \multicolumn{1}{c|}{7} & 0.19\% &  & 0.38\% &  & 3.35\% &  & 0.93 \\
        $\bm{\varepsilon}_\text{pl}$  & \multicolumn{1}{c|}{4} & 0.06\% &  & 0.20\% &  & 3.59\% &  & 0.99 \\
        \bottomrule
    \end{tabular}
\end{table}

\noindent
Despite minor differences, the multitask model excels in almost all metrics, except for Mean MAE\% in Von Mises stress and the maximum equivalent strain.

\begin{table}[h]
    \caption{Multi Task Improvements.}
    \label{tab:Single_vs_multi_2}
    \centering
    \begin{tabular}{lcccccc|c}
        \toprule
         & \multicolumn{6}{c}{$\text{MAE\%}$} & $R^2$ \\
        \cmidrule(lr){2-7} \cmidrule(lr){8-8}
         & \multicolumn{2}{|c}{Mean} & \multicolumn{2}{c}{Std} & \multicolumn{2}{c|}{Max} & Weighted \\

        \midrule
        \multicolumn{1}{c|}{$\bm{u}$ }                     & -0.04\% &  & -0.04\% &  & -0.09\% &  & 0 \\
        \multicolumn{1}{c|}{$\sigma_\text{vm}$}            & +0.04\% &  & -0.02\% &  & -0.13\% &  & -0.02 \\
        \multicolumn{1}{c|}{$\bar{\varepsilon}_\text{pl}$} & -0.07\% &  & -0.09\% &  & +0.94\% &  & +0.01 \\
        \multicolumn{1}{c|}{$\bm{\varepsilon}_\text{pl}$}  & -0.01\% &  & -0.04\% &  & -0.30\% &  & 0 \\
        \bottomrule
    \end{tabular}
\end{table}
\noindent
Training \added{a new multitask model}\chglabel{1.3f} with an additional variable like total equivalent strain ($\bar{\varepsilon}$) further boosts performance, as shown by the improvements in Table \ref{tab:extra_variable}. \added{This leads to an expanded output space in the last layer, $\mathbf{Y} \in \mathbb{R}^{(35+14) \times 10}$ where 14 represents the dimensionality of the reduced data used to capture the total equivalent strain.}\chglabel{1.1f}
\begin{table}[h]
\centering
    \caption{Enhancements of extra variable $\bar{\varepsilon}$.}
    \label{tab:extra_variable}
    \begin{tabular}{lccc|c}  
        \toprule
        \multicolumn{1}{c}{} & \multicolumn{3}{c}{{MAE\%}} & \multicolumn{1}{c}{$R^2$} \\
        \cmidrule(lr){2-4} \cmidrule(lr){5-5}
        \multicolumn{1}{c|}{} & Mean & Std & Max & Weighted \\
        \midrule
        \multicolumn{1}{c|}{$\bm{u}$}                           & -0.02\%          & -0.09\% & -1.55\% & 0        \\
        \multicolumn{1}{c|}{$\sigma_\text{vm}$}            & +0.02\%           & +0.04\% & +1.71\% & 0.00     \\
        \multicolumn{1}{c|}{$\bar{\varepsilon}_\text{pl}$} & -0.02\%          & -0.08\% & -1.65\%  & 0.00     \\
        \multicolumn{1}{c|}{$\bm{\varepsilon}_\text{pl}$}       & -0.02\%          & -0.05\% & -1.59\% & +0.01    \\
        \bottomrule
    \end{tabular}
\end{table}

\noindent
Incorporating $\bar{\varepsilon}$ enhances most variables but increases Von Mises stress error, likely due to a misalignment between this variable and $\bar{\varepsilon}$, affecting the shared representation's accuracy for Von Mises stress. Nevertheless, this representation improves accuracy for variables like equivalent plastic strain.

\begin{figure}[H]
  \centering
  \begin{subfigure}{0.49\textwidth}
    \centering
    \includegraphics[width=0.9\textwidth]{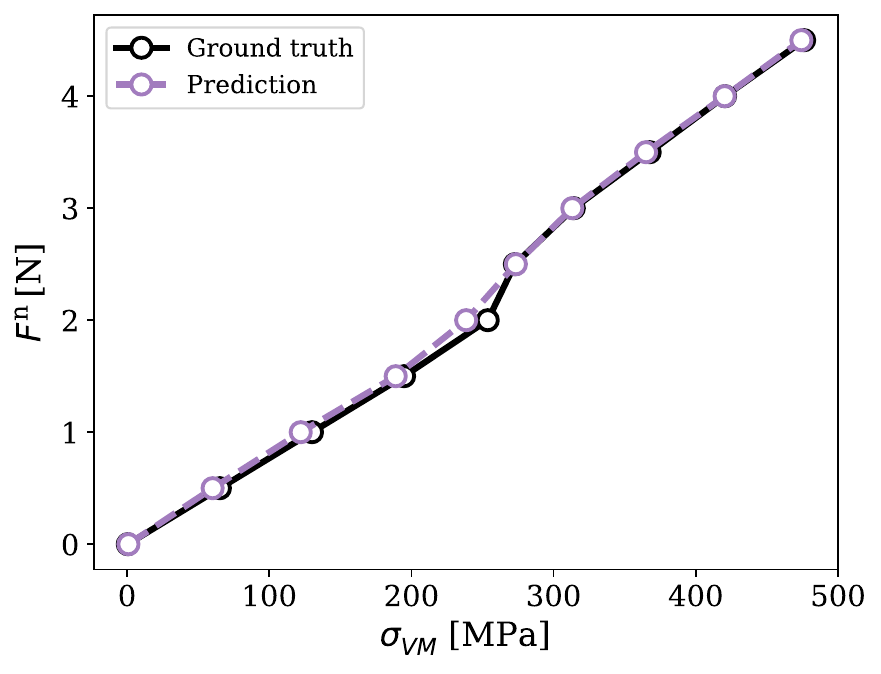}
    \caption{Sample 1: Force-stress}
    \label{fig:force_stress_1_2}
  \end{subfigure}
  \hfill
  \begin{subfigure}{0.49\textwidth}
    \centering
\includegraphics[width=0.9\textwidth]{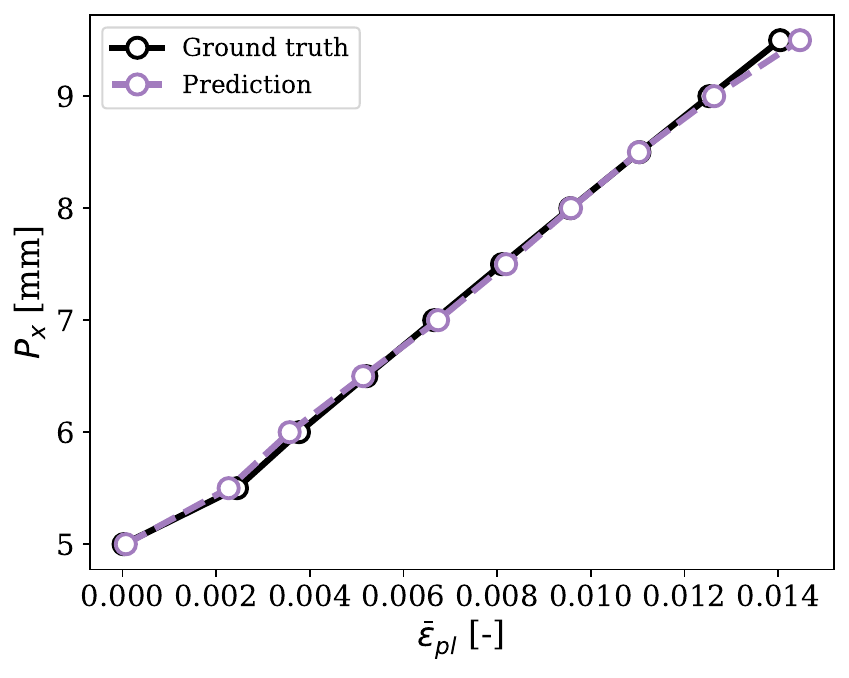}
    \caption{Sample 4: Position-plastic strain}
    \label{fig:position_strain_4_2}
  \end{subfigure}
  \caption{Effect of extra variable $\bar{\varepsilon}$ at point K.}
  \label{fig:beam_samples_2}
\end{figure}

These effects are shown in the updated Figure \ref{fig:beam_samples} in Figure \ref{fig:beam_samples_2}. Here, the discrepancy in the Von Mises stress relative to the ground truth has grown, whereas the equivalent strain now aligns more closely with the ground truth.

\begin{figure}[H]
  \centering
  \begin{subfigure}{0.49\textwidth}
    \centering
    \includegraphics[width=0.9\textwidth]{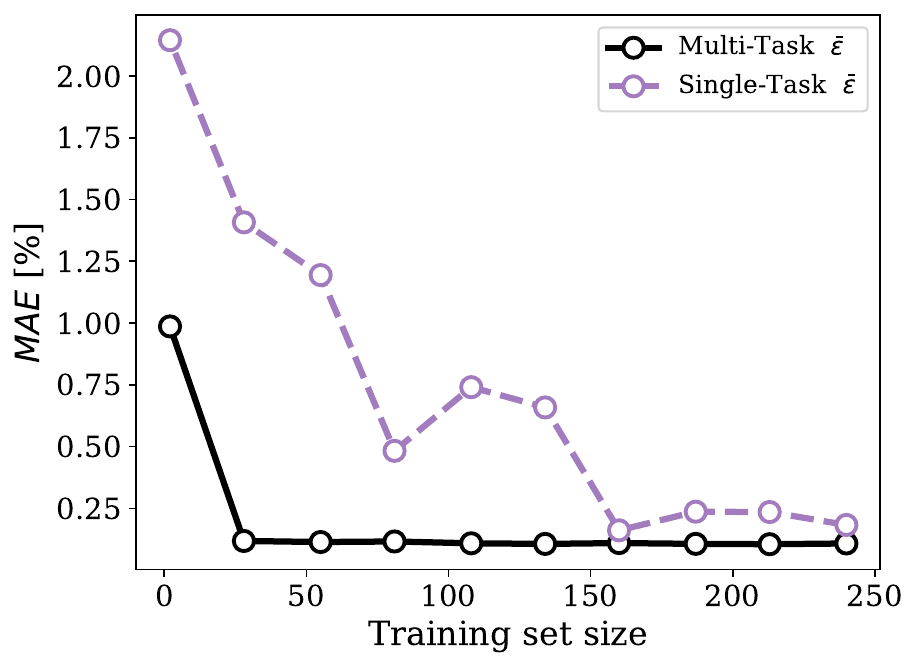}
    \caption{Mean absolute error}
    \label{fig:Sensitivity_eqtot_beam_single}
  \end{subfigure}
  \hfill
  \begin{subfigure}{0.49\textwidth}
    \centering
    \includegraphics[width=0.9\textwidth]{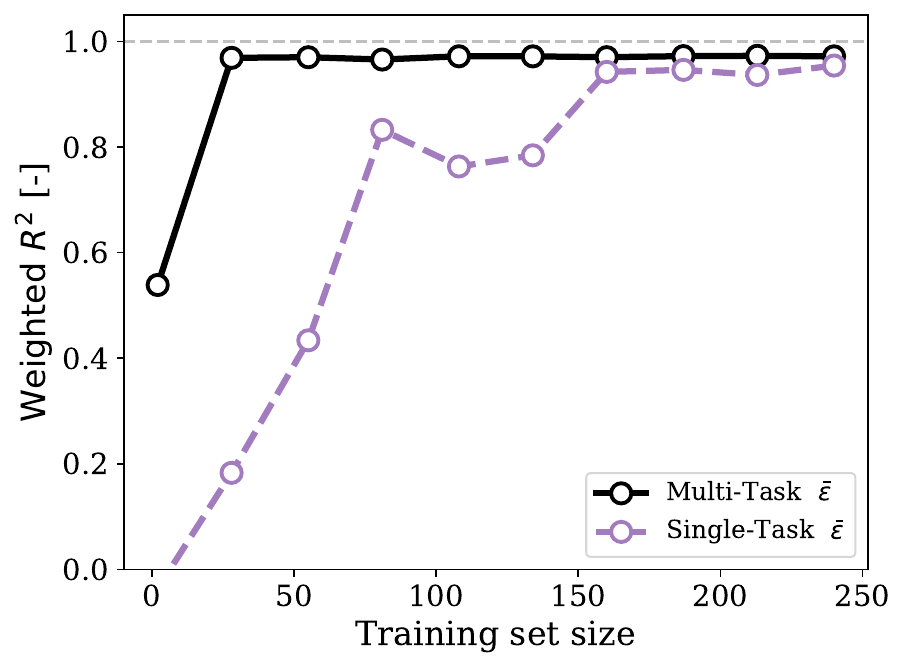}
    \caption{Weighted coefficient of determination}
    \label{fig:Sensitivity_eqtot_beam_multi}
  \end{subfigure}
  \caption{Test set: Model comparison on $\bar{\varepsilon}$ across different training set sizes.}
  \label{fig:eqtot_sensitivity}
\end{figure}

 \noindent
Additionally, a pre-trained multi-task model can effectively train on new variables with fewer samples. For example, training for equivalent strain \added{while keeping the other parameters fixed}\chglabel{1.3g}, converges with about 20 samples, in contrast to the 160 samples needed for the single-task model, as demonstrated in Figure \ref{fig:eqtot_sensitivity}.
\noindent
\subsection{Inference Time}
The networks' inference times and FEM computation times are detailed in Table \ref{tab:times} with hardware specifications.

\begin{table}[h]
\centering
\caption{Online computation time comparison.}
\label{tab:times}
\begin{tabular}{p{1.2cm}p{2.5cm}p{2.7cm}p{3.5cm}}
\toprule
 Use Case & Method & Time per Sample [s] & Hardware \\
\midrule
 1 \& 2 & Multi-Task LSTM & \num{6.8e-5} & NVIDIA T4 GPU\\
 1 & FEM & \num{1.6e2} & Intel Core i7-7700HQ CPU \\
 2 & FEM &  \num{2.0e2} & Intel Core i7-7700HQ CPU \\
\bottomrule
\end{tabular}
\end{table}

\noindent
The NN uses GPUs for faster parallel processing, while FEM relies on CPUs for sequential computations. This leads to the networks being over $2\times 10^6$ times faster than FEM, showing similar inference times in both use cases for uniform efficiency.
\added{
This comparison primarily illustrates computational demand, though it is somewhat imbalanced, given that PyFEM’s Python-based FEM is slower than optimized, GPU-based deep learning frameworks running on different hardware.}\chglabel{1.17a}

An important aspect of these ML models is their capability for real-time computation, crucial for applications requiring immediate data processing or rapid response. Real-time applications typically need at least 30 frames per second \citep{Brown2002AlgorithmicSimulation}, translating to a maximum of \num{3.33e-2} seconds per frame for each \replaced{load step}{time step}\chglabel{2.4gl}. Despite the lengthy initial data generation for training, the data reduction and training phases are quick, completing in under 2 minutes, highlighting the LSTM MTL framework's advantage in rapid processing.

\section{Conclusion}\label{sec5}
This study presents a novel LSTM multi-task framework for simulating elasto-plastic deformation, combining proper orthogonal decomposition and recurrent neural networks to efficiently deliver real-time 2D predictions. The model excels in predicting nonlinear, path-dependent behavior, achieving low error rates and high weighted $R^2$ values, effectively capturing complex material phenomena such as the Bauschinger effect. Its multi-task approach offers superior generalization and training efficiency over single-task models, thanks to shared layers that reduce overfitting. Moreover, a pre-trained multi-task model efficiently trains with fewer samples, demonstrating a comprehensive grasp of the task.

However, the research also identifies areas for improvement. LSTM models, while robust, are subject to compound errors, suggesting a need for increased trainable parameters to enhance performance. Furthermore, integrating extra variables to the MTL network improves most outcomes but can increase errors in others due to mismatches in shared representations.

Despite its strong performance in nonlinear path-dependent analysis, the model shows limitations in extrapolating near data boundaries and in scenarios with materials remaining in the plastic regime. Addressing these challenges involves incorporating additional variables to improve accuracy in complex scenarios with intricate geometries and loading conditions. Expanding tests to more sophisticated problems is anticipated to significantly enhance the model’s predictive capabilities, underscoring its applicability in advanced mechanical engineering applications.
\backmatter
\small
\bmhead{Acknowledgements}
The initial draft of this manuscript was composed by the authors, which was subsequently refined and enhanced with the aid of OpenAI's ChatGPT \citep{OpenAI2023ChatGPTVersion}.

\bmhead{Author contributions}
RS: Conceptualization, Data curation, Formal analysis, Investigation, Methodology, Software, Validation, Visualization, Writing - original draft.
JR: Conceptualization, Software, Code support, Methodology, Supervision, Writing - review \& editing.
OS: Project administration, Funding acquisition, Methodology, Supervision, Writing - review \& editing.
OM: Methodology, Writing - review \& editing.

\bmhead{Funding}
This work was partly funded by the Xecs TASTI project, number 2022005.

\bmhead{Availability of data and materials}
The datasets and source code used during the current study are available from the corresponding author on request.

\section*{Declarations}

\subsubsection*{Competing interests}
The authors declare that they have no competing interests.


\end{document}